\documentclass{aa}  
\usepackage{graphicx}
\usepackage{amsmath,amssymb} 
\usepackage{txfonts}
\usepackage[T1]{fontenc}
\usepackage{afterpage}
\usepackage{amssymb}
\usepackage{natbib} 
\usepackage{wrapfig}
\usepackage{graphicx}
\usepackage{soul}
\usepackage{color, colortbl}
\bibpunct{(}{)}{;}{a}{}{,}
\newcommand{\Ms}{\ensuremath{M_\odot}}
\newcommand{\Rs}{\ensuremath{R_\odot}}
\newcommand{\Mdot}{\ensuremath{\dot{M}}}
\newcommand{\ddr}[1]{\frac{\partial{}#1}{\partial{}r}}

 \newcommand{\diff}{{\rm d}}

 \newcommand{\lc}{ \left[}
 \newcommand{\rc}{ \right]}
 
 \def\lesssim{\mathrel{\hbox{\rlap{\hbox{\lower4pt\hbox{$\sim$}}}\hbox{$<$}}}}
 \def\gtrsim{\mathrel{\hbox{\rlap{\hbox{\lower4pt\hbox{$\sim$}}}\hbox{$>$}}}}
\usepackage{color}

\definecolor{Gray}{gray}{0.9}

\authorrunning{Amard et al.}
\titlerunning{Rotating models of young solar-type stars}
\begin{document}

      \title{Rotating models of young solar-type stars} %
 \subtitle{Exploring braking laws and angular momentum transport processes}

   \author{L. Amard\inst{1,2}
          \and 
          A. Palacios\inst{1}
          \and 
          C. Charbonnel\inst{2,3}
         \and
          F. Gallet\inst{2}
          \and 
          J. Bouvier\inst{4,5}
          }

   \offprints{A. Palacios or L. Amard}

   \institute{LUPM, Universit\'e de Montpellier, CNRS, Place
               E. Bataillon - cc 072, F-34095 Montpellier Cedex 05, France
                \\
                \email{louis.amard@umontpellier.fr}
                \and
               Department of Astronomy, University of Geneva, Chemin des Maillettes 51, CH-1290 Versoix, Switzerland
               \and 
                IRAP, UMR 5277, CNRS and Universit\'e de Toulouse, 14 Av. E. Belin, F-31400 Toulouse, France
        \and 
                Universit\'e Grenoble Alpes, IPAG, F-38000 Grenoble, France
                \and 
                CNRS, IPAG, F-38000 Grenoble, France\\
             }

   \date{Received; accepted}

   \abstract{Understanding the angular momentum evolution of stars is one of the greatest challenges of modern stellar physics.}%
   {We study the predicted rotational evolution of solar-type stars from the pre-main sequence to the solar age with 1D rotating evolutionary models including physical ingredients.}%
   {We computed rotating evolution models of solar-type stars including an external stellar wind torque and internal transport of angular momentum following the method of   \textup{Maeder and Zahn} with the code STAREVOL. We explored different formalisms and prescriptions available from the literature. We tested the predictions of the models against recent rotational period data from extensive photometric surveys, lithium abundances of solar-mass stars in young clusters, and the helioseismic rotation profile of the Sun.}%
   {We find a best-matching combination of prescriptions for both internal transport and surface extraction of angular momentum. This combination provides a very good fit to the observed evolution of rotational periods for solar-type stars from early evolution to the age of the Sun. Additionally, we show that fast rotators experience a stronger coupling between their radiative region and the convective envelope. Regardless of the set of prescriptions, however, we cannot simultaneously reproduce
surface angular velocity and the internal profile of the Sun or the evolution of lithium abundance. }%
   {We confirm the idea that additional transport mechanisms must occur in solar-type stars until they reach the age of the Sun. Whether these processes are the same as those needed to explain recent asteroseismic data in more advanced evolutionary phases is still an open question.}%


\maketitle
%

\section{Introduction}

Stars are born out of collapsing molecular clouds, which determine their masses, initial chemical composition, and initial angular momentum (hereafter AM) content. It has become clear in the past few decades that rotation can strongly influence the stellar evolution \citep{ES76,MaederMeynetARAA2000,Maederbook}, which
makes understanding the rotational evolution of stars one of the greatest challenges in stellar physics.\\
Particular interest has been given to the rotational evolution of solar-mass stars. Photometric surveys of open clusters and stellar associations \citep[see e.g.][for a recent review]{BouvierPPVI} provide a wealth of rotation periods from ages of 1 Myr to the age of the Sun. The patterns revealed by the rotational periods indicate that a large dispersion exists among stars in the mass range 0.9-1.1 M$_\odot$ during the pre-main sequence (PMS) that narrows as they evolve, and converges towards a single peaked distribution by the time they reach an age of about 1 Gyr. The main trends of the rotational evolution of solar-type stars through the PMS, zero-age main sequence (ZAMS), and main sequence (MS) can be described by considering three types of initial rotation, as suggested by statistical analyses of the observed samples \citep{IrwinBouvier2009,GB13}: fast, medium, and slow rotation. \\ Several models have been built to account for the observed evolution of surface velocity \citep[e.g.][]{ES81,MB1991,KPBS97,MacGregor2000,Denissenkov2010,Spadaetal2011,Charbonnel2013,GB15,LS15}, which all include three fundamental processes: (1) a star-disc interaction phase, often modelled as a constant surface angular velocity phase,
(2) a coupling mechanism between the radiative interior and the convective envelope that regulates the angular momentum exchange between these two reservoirs during the evolution, and (3) a braking torque expression that is due to the loss of angular momentum extracted by the stellar winds and leads to a decrease of the surface angular velocity mostly during the MS evolution. All these studies have highlighted that the torque and the internal coupling processes are important for reproducing the observations.

In the present paper, we extensively test the different prescriptions available to date for both stellar wind torque and the turbulence modelling within the self-consistent theoretical framework provided by \citet{Zahn1992} and \cite{MaederZahn1998} to treat the internal transport of angular momentum by meridional circulation and shear-induced turbulence in 1.5D \citep[and reference therein]{DecressinMathis2009}. These later internal processes successfully account in a consistent way for the rotational behaviour and the evolution of the surface abundances of light elements for dwarf and subgiant stars with masses above $\sim$ 1.4 M$_\odot$ \citep[the blue side of the so-called lithium dip;][]{CT99,PTCF2003}. However, their efficiency in transporting AM has been shown to be too low in solar-type stars to reproduce the present-day solar internal rotation as inferred from helioseismic data and the Li abundance in solar type stars \citep{TalonCharbonnel1998}; for these objects, additional AM transport processes may be required to better fit the data, such as internal gravity waves or magnetic fields \citep{CharbonnelTalon2005Science,Eggenberger2005}. Despite this assessment concerning lithium and helioseismology, AM extraction and rotation-induced mechanisms have become the basic processes upon which more elaborate models can be built \citep{TalonCharbonnel2003,Lagarde2012,Charbonnel2013}. It is thus important to clarify their exact efficiency and also, if possible, the more appropriate combination of prescriptions that should be used to model solar-mass stars, in a similar way as \citet{Meynet2013} did for the case of massive stars.
Unlike massive stars, AM transport mechanism in solar-like stars are driven by the surface extraction of AM. New magnetic torque prescriptions have recently been developed and have to be accounted for to find a more appropriate combination of prescriptions.

We present the different formalisms that we test in Sect.~\ref{sec:formalism} and recall the overall physics of the models in Sect.~\ref{sec:models}. Next, we describe our reference model in Sect.~\ref{sec:ref} and then explore the parameter space in Sect.~\ref{sec:explore}. In Sect.~\ref{sec:bizone} we compare our results to the recent bi-zone models by \citet{GB13,GB15}.  Finally, we briefly compare our models to the additional constraints given by helioseismology and the surface abundances of lithium in Sect.~\ref{sec:other}. A summary of our results and some perspectives are given in Sect.~\ref{sec:end}.

\section{Formalism}\label{sec:formalism}

\subsection{Disc coupling}
A solar-mass star undergoes a global contraction during the first few million years of its evolution. During this period the star magnetically interacts with its circumstellar accretion disc, and this interaction modifies the stellar angular momentum. 
The final stellar velocity results from the balance between the increase of AM from the accreted matter and the strong loss that can be due either to accretion-powered stellar winds \citep{MattPudritz2005} or to the so-called disc-locking process\citep{Koenigl1991,GL79}.

While these mechanisms still lack an accepted physical description, it is observationally evident \citep[e.g.][and references therein]{Rebull2004} that this interaction is very effective during the whole disc lifetime and compensates for the increase of angular velocity that is due to the stellar contraction. 
In this configuration, the star will, on average, maintain a constant angular velocity as long as it is coupled with the disc. This is confirmed by the analysis of rotation periods in young open clusters and associations \citep{GB13}.

The duration of this period of coupling varies with both the mass \citep{KennedyKenyon2009} and the initial angular momentum of the star.
To evaluate the duration of this strong star-disc interaction phase, it is necessary to rely on observations since there is a degeneracy between the initial angular momentum content of the star and the disc-coupling time. A star with a large initial angular momentum content that will maintain a coupling with its circumstellar disc for a relatively long time is expected to undergo the same rotational evolution for the remaining evolution as a star with a smaller initial angular momentum content coupled to a shorter lived disc. \citet{Edwardsetal1993} showed that the observed rotation periods of stars in very young clusters demand shorter disc lifetimes for fast rotators than for slow rotators. This has been confirmed with more observational evidence by \citet{Rebull2004} and \citet{Bouvier2008}.

In the present study, we assume a 2.5 Myr and 5 Myr star-disc interaction for fast and medium or slow rotators, respectively, based on observations \citep[e.g.][]{Belletal2013}. Throughout this coupling, the surface rotation period is assumed to be constant.

\subsection{Stellar wind torque}\label{sec:torques}

\begin{figure}
\includegraphics[width=0.45\textwidth]{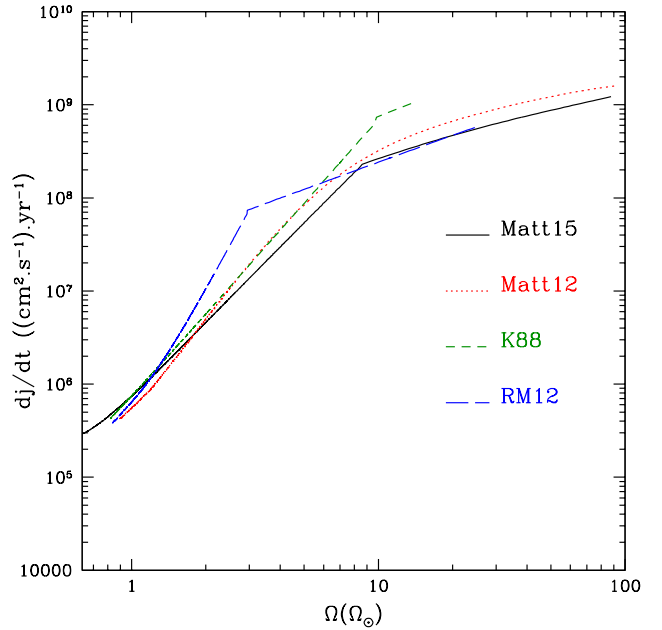}
\caption{Stellar wind torque as a function of the surface angular velocity for the different prescriptions as labelled in the plot. The parameter values used in each case are given in Table.~\ref{tab:brake}.}
\label{Fig:torqueomega}
\end{figure}
 For the past two decades, many different more or less complex
prescriptions of stellar wind braking have been developed. 
Since \cite{Schatzman1962} and \citet{WD67}, all the processes playing a role in the coupling between wind-driven mass-loss and the magnetic field have been extensively studied and improved.
In a general manner, the net torque exerted on the star can be written as  
\begin{equation}
\frac{{\rm d}J}{{\rm d}t} = \Mdot \Omega_\star r_A^2
,\end{equation}
where $J$ is the total angular momentum of the star, $\Mdot$ is the mass-loss rate, $\Omega_\star$ the average angular velocity, and $r_A$ the average Alfven radius. The most difficult term to constrain in this equation,  which is also the most important one, is the Alfven radius, which depends on several parameters such as the stellar radius, effective temperature, and magnetic field.\\
In the following, we present the different prescriptions we used to evaluate their impact on the evolution of the surface velocity and internal rotation profile of a single 1\Ms~star. All the prescriptions detailed below are semi-empirical, based on theoretical and observational properties. Except for the expression from \cite{RM12} (detailed in part \ref{RM12}), they all lead to a rotational evolution that follows the empirical relationship
of \cite{Skumanich72},   $\Omega_S \propto t^{-1/2}$ , beyond the spin-down phase of the model stars (after at least 1 Gyr of evolution). We calibrated the constants of each braking law so that it leads to the solar rotation rate at the age of the Sun for a 1~\Ms~model with our reference prescription. The constants are summarised in Table~\ref{tab:brake}. Figure~\ref{Fig:torqueomega} shows the torque evolution as a function of the surface angular velocity on the main sequence for each of the prescriptions.

\subsubsection{Kawaler (1988) - Chaboyer et al. (1995a)}\label{K88}
 \cite{Kawaler88} derived a braking law that takes into account the magnetic field strength and topology and the mass-loss rate. The expression can easily be incorporated in stellar evolutionary codes accounting for rotation \citep[e.g.][]{Pinsonneaultetal1989}
 
 \begin{equation}
\frac{dJ}{dt} = -K_W\Omega^{1+(4an/3)}\left(\frac{R}{\Rs}\right)^{2-n}\left(\frac{\Mdot}{-10^{-14}\Ms/yr^{-1}}\right)^{1-(2n/3)}\left(\frac{M}{\Ms}\right)^{-n/3}.
\end{equation}

Here, the angular momentum loss is directly related to the rotation rate $\Omega$. It only depends on three parameters: the exponent $a$ that depicts the total magnetic field strength dependence on the rotation rate ($B_0 = K_B\left(R/\Rs\right)^{-2}\Omega^a$ , with $B_0$ the surface magnetic field strength and $K_B$ constant) fixed equal to $1$; the wind-index factor $n$ that varies with the magnetic field topology and is typically assumed to be equal to $3/2$ to reproduce the Skumanich relationship; and the parameter $K_W$. The latter consists, in the initial paper, of two components linked on one hand to the dependence of the magnetic field generation to the convection zone depth and on the other hand to the structure of the stellar wind. It is here assumed to be a single constant that is calibrated to reproduce the solar surface velocity at the age of the Sun.
Taking $n = 3/2$ simplifies the problem by suppressing the dependence
on mass loss rate, leading to 
\begin{equation}
\frac{dJ}{dt} = -K_W\Omega^{(1+2a)}\left(\frac{R}{\Rs}\right)^{1/2}\left(\frac{M}{\Ms}\right)^{-1/2},
\label{Eq:K88a}
\end{equation}
where $K_W = 2.10^{48}$ is calibrated in cgs units to reproduce the solar case, and $a = 0$ or $1$ for a saturated or unsaturated magnetic field, respectively.

Following \cite{StaufferHartmann87}, \cite{Chaboyeretal1995a} assumed that the magnetic field reaches saturation above a certain value of the surface angular velocity. As a result, this saturation modifies the braking law and the dependence of the angular momentum loss on the rotation rate.
It is not yet clear whether this saturation is due to the saturation of the dynamo itself or to coronal processes \citep{Wright2011}. However, the magnetic field strength seems to stop increasing even if the star still spins up \citep{Vilhu84,Odelletal95} and $a = 0$ in the expression of $B_0$.
Equation \ref{Eq:K88a} then becomes 
\begin{equation}
\frac{dJ}{dt} = -K_W\Omega\Omega_{sat}^2\left(\frac{R}{\Rs}\right)^{1/2}\left(\frac{M}{\Ms}\right)^{-1/2} \qquad {\rm for} \; \Omega \geq \Omega_{sat}
\end{equation}
\begin{equation}
\frac{dJ}{dt} = -K_W\Omega^{3}\left(\frac{R}{\Rs}\right)^{1/2}\left(\frac{M}{\Ms}\right)^{-1/2} \qquad  {\rm for} \; \Omega < \Omega_{sat}
,\end{equation}
which is the formulation that we use in our stellar evolution code.
\subsubsection{\cite{Matt2012}}\label{Matt12}

\cite{Matt2012} proposed a braking law based on current 2D magnetohydrodynamical (MHD) simulations for stellar winds. 
They varied the magnetic field strength relative to the mass-loss rate and surface gravity to derive the most complete stellar wind torque formula available at the time for solar-type stars,

\begin{equation}
\frac{dJ}{dt} = - \frac{K_1^2}{\left(2G\right)^m}\bar{B}^{4m}\Mdot^{1-2m}_W\frac{R^{5m+2}}{M^m}\frac{\Omega}{\left(K^2_2+0.5u^2\right)^m},
\label{eq:Matt12}
\end{equation}
with $u$ being the ratio of the surface velocity to the brake-up velocity, and $K_1$, $K_2$ and $m$ tunable (to a certain point) parameters to fit the observations. The adopted values were taken from \citet{GB13} and are given in Table~\ref{tab:brake}.

\begin{table}
\caption{Parameters used for the prescriptions from \cite{Kawaler88} and \cite{Chaboyeretal1995a} ({\sf K88}), \cite{Matt2012} ({\sf Matt12}), \cite{Mattetal2015} ({\sf Matt15}), and \cite{RM12} ({\sf RM12}).}
\begin{tabular}{ c | c | c | c }
\hline
\hline
{\sf K88}& {\sf Matt12} & {\sf Matt15} & {\sf RM12}\\
\hline
$K_W = 2.10^{48} $ & $K_1 =6.7$ & $K = 5.10^{31}$ & $\mathcal{C} = 10^{39}$ \\ 
$n = 3/2$ &  $K_2 = 0.506$ & $m = 0.22$ & $\Omega_{sat} = 3\Omega_\odot$ \\ 
$\Omega_{sat} = 10 \Omega_\odot$ & $m = 0.17$ & $p = 1.7$ \\ 
& & $\chi = 10$\\ 
\hline
\end{tabular}
\label{tab:brake}
\end{table}

We directly included a modified version of the BOREAS subroutine \citep{CS2011,GB13} in our evolution code STAREVOL (see the Appendix for details concerning the applied modifications to the original distributed version of BOREAS) to obtain the mean magnetic field $\bar{B} = f_* B_*$ at each time step, $f_*$ being the filling factor expressing the magnetized fraction of the stellar surface \citep{Saar1996}.
We had to recalibrate the expression of $f_*$  to reach the solar mass-loss value at the age of the Sun for solid-body rotating models,
\begin{equation}
f_\star = \frac{0.4}{[1+\left(x/0.16\right)^{2.3}]^{1.22}},\end{equation}
with $x$ being the normalised Rossby number\footnote{$\tau_{conv}$ and $\tau_{conv_\odot}$ are determined as a function of the effective temperature as in \citet{CS2011}.} $x = \frac{P_{rot}}{P_{rot_\odot}}\frac{\tau_{conv_\odot}}{\tau_{conv}}$.

With this expression, the magnetic field reaches a saturation threshold for $\Omega_{sat} \approx 10 - 15 \Omega_\odot$.\\ 
We also used the mass loss accounting for magnetohydrodynamic turbulence and Alfv\'en waves that is an output of the BOREAS routine. We accounted for mass loss from the end of the disc-coupling phase during the pre-main sequence to the end of the main sequence considering that the mechanisms responsible for mass loss remain the same during these phases. 

As shown by \cite{GB13}, Eq.~\ref{eq:Matt12} combined with the prescriptions of \cite{CS2011} as computed with the modified BOREAS routine allows reproducing the observed rotation periods well in the validity domain of \citet{CS2011}'s work, which is in particular the case of stars of about 1~\Ms. \cite{VSP2013} extended the application to more massive stars (1.1-1.5\Ms) by combining this prescription with the mass-loss rate of \citet{Woodetal2005}.

\subsubsection{\cite{RM12}}\label{RM12}
\citet{RM12} introduced a formalism in which the  magnetic field \textit{strength} is accounted for as directly depending on the Rossby number in the unsaturated regime (while it is the magnetic \textit{flux} that depends on the Rossby number in the approach of \citet{Matt2012}), 
\begin{equation}
\frac{dJ}{dt} = -\mathcal{C} \left(\frac{R^{16}}{M^2}\right)^{1/3} \Omega \qquad {\rm for }\; \Omega \geq \Omega_{sat}, 
\end{equation}
\begin{equation}
\frac{dJ}{dt} = -\mathcal{C}  \left(\frac{R^{16}}{M^2}\right)^{1/3} \left(\frac{\Omega}{\Omega_{sat}}\right)^4 \Omega \qquad {\rm for }\; \Omega < \Omega_{sat},
\label{eq:RM12}
\end{equation}
with
\begin{equation}
\mathcal{C}=\frac{2}{3}\left(\frac{B^8_{crit}}{G^2K_V^4\Mdot}\right)^{1/3} \nonumber
\end{equation}
being a constant because each parameter on the right-hand side is assumed to adjust to keep the overall product constant throughout the evolution.

The main consequence of this change of paradigm is a torque that strongly depends on the stellar radius (${\rm d}J/{\rm d}t\propto R^{16/3}$!) and hence on the stellar mass and evolutionary stage. As shown in Fig. \ref{Fig:torqueomega}, the saturation value is reached at a very low rotation rate, which drastically changes the behaviour of surface rotation rate. The extraction of AM can be one order of magnitude higher for rotation rates between 1 and 10 times the solar value with this braking law.

This prescription shows some important weaknesses in reproducing the rotation rates of stars in open clusters as a function of mass, as shown by \citet[][their Fig.4]{RM12}. At the age of the Hyades, they obtain in particular a decrease of $P_{rot}$ with decreasing mass for M $\ge 0.3$ M$_\odot$, which contradicts observations. Reiners and Mohanty\textcolor[rgb]{1,0.501961,0}{} interpreted this discrepancy as the result of a possible core-envelope decoupling that would be proportional to the size of the radiative core (which increases with increasing mass at a given age). To test this hypothesis, we used this prescription for the stellar wind torque in our self-consistent differentially rotating 1~\Ms~models.

\subsubsection{\cite{Mattetal2015}}
The main improvement reached by \citet{Mattetal2015} over  \cite{Matt2012} is to highlight the Rossby number as a critical parameter, which was first reported by \citet{Soderblom85} and \citet{Vilhu86}. Compared to \citet{Matt2012}, \cite{Mattetal2015} expressed the torque as a simple function of mass, radius, angular velocity, and turnover timescale of the convective envelope\footnote{Determined as a function of T$_{eff}$ as in \citet{CS2011}.}.
Starting from a general expression of the torque from \cite{Matt2012}, 
\begin{equation}
\frac{dJ}{dt} = \left(\frac{dJ}{dt}\right)_\odot \left(\frac{M_\star}{M_\odot}\right)^{-m} \left(\frac{R_\star}{R_\odot}\right)^{5m+2} \left(\frac{B_\star}{B_\odot}\right)^{4m} \left(\frac{\Mdot_\star}{\Mdot_\odot}\right)^{1-2m} \left(\frac{\Omega_\star}{\Omega_\odot}\right),
\end{equation}
they expressed the more uncertain terms, namely $\Mdot_\star$ and $B_\star$, as a function of the Rossby number. This results in some equations according to the regime of the magnetic field (saturated or not),
\begin{equation}
\frac{dJ}{dt} = -\mathcal{T}_0 \left(\frac{\tau_{CZ}}{\tau_{CZ\odot}}\right)^{p} \left(\frac{\Omega_\star}{\Omega_\odot}\right)^{p+1}  \rightarrow {\rm unsaturated},
\end{equation}
\begin{equation}
\frac{dJ}{dt} = -\mathcal{T}_0 \chi^p \left(\frac{\Omega_\star}{\Omega_\odot}\right)  \rightarrow  {\rm saturated},
\end{equation}
with 
\begin{equation}
\mathcal{T}_0 = K \left(\frac{R_\star}{R_\odot}\right)^{3.1} \left(\frac{M_\star}{M_\odot}\right)^{0.5}\gamma^{2m}  ,
\end{equation}
where $\gamma = \sqrt{1+(u/0.072)^2}$ comes from Eq. $(8)$ of \cite{Matt2012}, and $u$ has the same meaning as above. $\chi = \frac{Ro_\odot}{Ro_{\rm sat}}$ is a constant. Considering that for $Ro \leq Ro_{\rm sat}$ the magnetic activity stops increasing and saturates at an approximately constant value. 
The calibrated values of the parameters $K,p,m$ and $\chi$ are given in Table~\ref{tab:brake}.

\subsection{Transport of angular momentum in the stellar interior}
\subsubsection{Global equations}

We treated the angular momentum evolution from the first model on the PMS following the formalism developed by \citet{Zahn1992} and \citet{MaederZahn1998}. This formalism assumes a strong anisotropy in turbulence, the horizontal turbulence (i.e. along isobars) being much stronger than the vertical one (i.e. perpendicular to the isobars), thus enforcing a shellular rotation. The transport of AM in the stellar interior follows the advection/diffusion equation,
\begin{equation}
\rho \frac{{\rm d}}{{\rm d}t}\left(r^2\Omega\right)= \frac{1}{5r^2}\ddr{} \left(\rho r^4 \Omega U_r\right) + \frac{1}{r} \ddr{} \left(r^4\rho \nu_v \ddr{\Omega}\right),
\label{eq:general}
\end{equation}
where $\rho$, $ r$, $\nu_v$ and $U_r$ are the density, radius, vertical component of the turbulent viscosity, and the meridional circulation velocity on a given isobar, respectively. 

This equation applies to radiative regions, while convective zones are assumed to rotate as solid bodies. The torque discussed in the previous section is applied at the upper convective boundary \citep[see][]{PTCF2003},

\begin{displaymath}
\begin{array}{ll}
{\displaystyle \frac{\partial}{\partial t}
\lc \Omega \int _{r_t}^R  r^4 \rho \, \diff r \rc =
-\frac{1}{5} r^4 \rho \Omega U + {\cal F}_\Omega } & ~~~{\rm for} ~ r=r_t \vspace{0.2cm}\\
\end{array}
\end{displaymath}
where ${\cal F}_\Omega$ is the torque, and $r_t$ is the radius at the lower edge of the convective envelope.\\

By integrating the angular momentum transport Eq. (\ref{eq:general}) over the surface of radius $r_{cz}$, $r_{cz}$ being the radius of the inner convective zone boundary, we obtain a flux equation,
\begin{equation}
F_{\rm tot} = F_{S}(r_{\rm cz}) + F_{MC}(r_{\rm cz})
\label{eq:fluxeq}
,\end{equation}
with
\begin{equation}
F_{S}(r_{\rm cz}) = \frac{\rm{d} J_{S}}{\rm{d} t} \bigg|_{r=r_{cz}} = -\rho r^4 \nu_v \ddr{\Omega}\bigg|_{r=r_{cz}}
\end{equation}
the flux carried by shear-induced turbulence from the radiative zone to the convective envelope, and
\begin{equation}
F_{MC}(r_{\rm cz}) = \frac{\rm{d} J_{MC}}{\rm{d} t} \bigg|_{r=r_{cz}}= -\frac{1}{5} \rho r_{cz}^4 \Omega U_{r_{cz}}
\end{equation}
the flux carried by meridional circulation. A detailed derivation of the AM fluxes is given in
 \cite{DecressinMathis2009} as part of a set of tools for assessing the relative importance of the processes involved in AM transport in stellar radiative interiors.

\begin{table}[h]
\caption{Different prescriptions used for the turbulent diffusion coefficients in our models.}
\begin{center}
\begin{tabular}{| c | c | c }
\hline\hline
\rowcolor{Gray}Prescription & $D_h \equiv \nu_h$\\ \hline\hline
& \\
\citet{MPZ2004}  & $r\sqrt{\left[Cr\overline\Omega \vert 2 V_r-\alpha U_r\vert\right]}$\\ 
({\sf MPZ04})& with $ \alpha={\frac{1}{2}}\,\frac{{\partial}\ln(r^2\overline\Omega)}{{\partial}\ln r}$ \\
& and $C = 1.6\times10^{-6}$ \\ 
& \\\hline
& \\
\citet{Zahn1992}  & $\frac{r}{C_{h}}\vert2 V_r-\alpha U_r\vert$\\ 
({\sf Zahn92})& with $C_h$ = 1\\
& \\\hline \hline
\rowcolor{Gray}Prescription & $D_v \equiv \nu_v$\\  \hline\hline
& \\
\citet{TZ97} & $\frac{Ri_{\rm c}}{N^2_{T} /(K_{T} + D_{h}) + N^{2}_{\mu}/ D_{h}} \left(r\frac{\partial \overline\Omega}{\partial r}\right)^2\,$\\
({\sf TZ97})& with $Ri_c = 0.25$\\
& \\ \hline
& \\
\citet{Maeder1997}  & $f_\text{energ} \frac{H_P}{g\delta}\frac{K}{\left[\frac{\varphi}{\delta}\nabla_\mu 
        + \left( \nabla_\text{ad} - \nabla_\text{rad} \right)\right]} 
    \left( \frac{9\pi}{32}\ \Omega\ \frac{\text{d} \ln \Omega}{\text{d} \ln r} \right)^2$\\
({\sf Maeder97})    & with $K = \frac{4ac}{3\kappa}\frac{T^4\nabla_\text{ad}}{ \rho P \delta}$\\
    & $f_\text{energ} = 1$ and $\varphi = \left( \frac{\text{d}\ln\rho}{\text{d}\ln\mu} \right)_{P,T} = 1$\\
\hline
\end{tabular}
\end{center}
\label{tab:coeffdiff}
\end{table}

Equation~\ref{eq:general} is complemented by the evolution equation for the relative mean molecular weight variations over an isobar $\Lambda = \tilde{\mu}/\bar{\mu}$ ,

\begin{equation}
\frac{{\rm d}\Lambda}{{\rm d}t} - \frac{{\rm d} \ln {\overline \mu}}{{\rm d}t} \Lambda = \frac{N_{\mu}^{2}}{{\overline g}\varphi}U_r-\frac{6}{r^2}\nu_h\Lambda\, , 
\label{Lambda}
\end{equation}
 where $\overline \mu$ is the mean molecular weight over an isobar, $\nu_h$ is the diffusion coefficient associated with the horizontal shear instability, and $N_{\mu}$, the chemical part of the  Brunt-V\"ais\"al\"a
  frequency is given by $N_{\mu}^2=\left({\overline
      g}\varphi/H_{P}\right)\nabla_{\mu}$ with $\nabla_{\mu}={\partial \ln 
    \overline{\mu}}/{\partial \ln P}$. \\

\subsubsection{Turbulence modelling}

As in \citet[e.g.][]{Zahn1992}, the shear turbulence in the horizontal and vertical directions is represented as a diffusive process. We assumed that the diffusion coefficients are well represented by the viscosities, that is, $D_v \approx \nu_v$ and $D_h \approx \nu_h$.\\ Several prescriptions exist for these viscosities, and we explored the effect of four different prescriptions on the angular momentum evolution of a  1\Ms~model. These prescriptions are summarised in Table~\ref{tab:coeffdiff}, and we refer to the associated papers for more details on how they were derived.\\
For the vertical diffusion coefficient, the prescription of \citet{TZ97} has been used for all the rotating models computed with the STAREVOL code so far \citep{PTCF2003,PalaciosCharbonnel2006,DecressinMathis2009,Lagarde2012}, while the prescription of \citet{Maeder1997} has been systematically used for the rotating models computed with the Geneva code \citep{Eggenbergeretal12,Georgy2013}. \citet{Meynet2013} have shown that the choice of the turbulent transport prescriptions may dramatically affect predictions for the structural, chemical, and rotational evolution of massive stars, but they were unable to clearly determine a best combination
for fitting the observational chemical constraints. In Sect.~\ref{sec:explore} we discuss this in a similar way within the framework of the rotational evolution of solar-type stars.

\subsection{Transport of chemicals}
The transport of chemical species in radiative region is computed as a purely diffusive process \citep{ChaboyerZahn92} but accounts for vertical advection and a strong horizontal diffusion. For a chemical species $i$, the concentration $c_i$ follows the equation 
\begin{equation}
\frac{{\rm d} c_i}{{\rm d}t} = \dot{c}_i + \frac{1}{\rho r^2}\ddr{}\left(\rho r^2 D_{tot} \ddr{c_i}\right),
\end{equation}
with $D_{tot} = D_{eff}+D_v$ the total diffusion coefficient, $D_{eff}$ given by 
\begin{equation}
D_{eff} = \frac{\vert rU(r)\vert^2}{30D_h},
\end{equation}
where $D_v$ and $D_h$ were defined previously (see Table \ref{tab:coeffdiff}). Finally, the term $\dot{c}_i$ refers to the temporal evolution of the concentration of chemical species $i$ due to nuclear burning. 

\begin{figure}[t]
\includegraphics[width=0.45\textwidth]{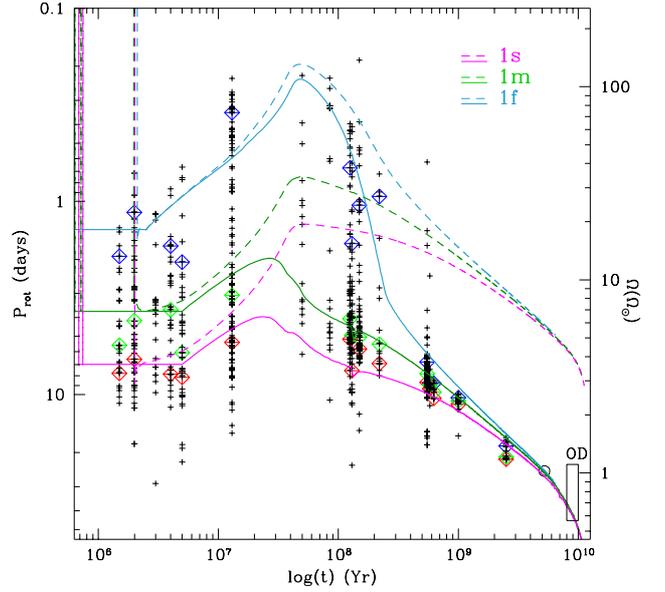}
\caption{Evolution of the angular velocity as a function of time for the reference models. Plotted on the left as the rotational periods in days and during the right as the angular velocity in solar units with $\Omega_\odot = 2.86.10^{-6}$s$^{-1}$. Each cross represents a measurement for a star belonging to an open cluster whose age has been taken from the literature. All observational data are taken from \citet[][and references therein]{GB15}  except the "OD" black frame between 8 and 10 Gyr, which corresponds to photometric data for stars observed with the {\sf KEPLER} satellite that belong to the disc of the Milky Way \citep{McQuillan2013}.
Solid lines show the evolution of surface rotation for the slow (magenta), medium (green), and fast (blue) rotating models, while short dashed lines represent the associated averaged angular velocity of the radiative interior.}
\label{Fig:refprot}
\end{figure}

\section{Physics of the models}\label{sec:models}

For all the models reported in Table~\ref{tab:mods}, the basic input physics (equation of state, nuclear reaction, opacities) can be found in \cite{Lagarde2012}. The initial abundances and mixing length parameter are calibrated without microscopic diffusion to reproduce a non-rotating Sun with respect to the \cite{AsplundGrevesse2009} solar mixture with a $10^{-5}$ precision for luminosity and radius at the age of the Sun. We used a mixing length parameter $\alpha_c = 1.7020$, an initial helium abundance $ Y = 0.2689 $, and an initial metal abundance $ Z = 0.0134 $. These values differ slightly from \citet{Lagarde2012} because we included a non-grey atmosphere treatment based on \cite{KrishnaSwamy}, which better agrees with observations for Sun-like stars \citep[e.g.][]{Vandenberg2007}.

In addition to the treatment of transport and loss of angular momentum and mass as described in Sect.~2, we included the modification of the effective gravity by the centrifugal forces and its effect on the stellar structure equations following \citet{ES76}. This effect is slightly visible on the evolutionary track in the Hertzsprung-Russell diagram when the stars rotate fast. 
Rotation can have a non-negligible impact on the effective temperature for the fastest rotators around the ZAMS. 

\begin{figure*}\includegraphics[width=0.95\textwidth]{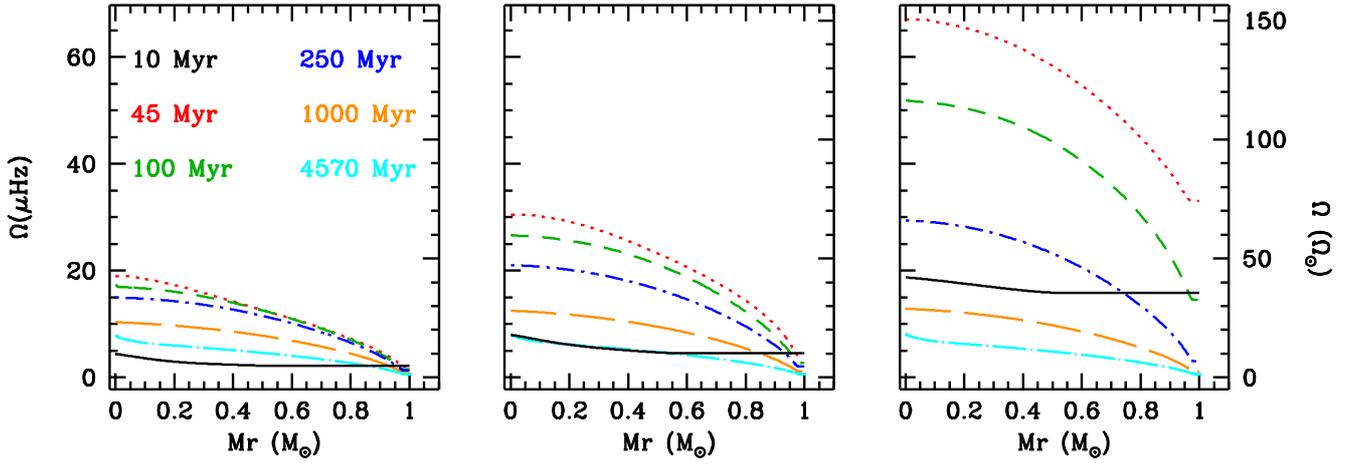}
\caption{Angular velocity profile of the reference models at different epochs (solid: 10 Myr; dotted: 45 Myr; dashed: 100 Myr; dot-dashed: 250 Myr; long-dashed: 1 Gyr; long-dash-dotted: 4.57 Gyr)  as a function of the relative mass fraction. The left, central, and right plots correspond to cases with slow, medium, and fast rotation (models $1s$, $1m$, and $1f$).}
\label{Fig:refprofomega}
\end{figure*}

\section{Reference model}\label{sec:ref}

The statistical analysis of the distributions of rotation periods in open clusters and associations from 1 Myr to 2.5 Gyr performed by \citet{GB13,GB15} focuses on three types of rotators: the slow, medium, and fast rotators, associated to the 25th, 50th,
and 90th percentiles of the statistical sample in each of the clusters. We focused on the same types of rotators and defined our reference models as those that reproduce their evolutionary paths, as shown in Fig.~\ref{Fig:refprot}.
They are labelled $1s$, $1m,$ and $1f$ in Table~\ref{tab:mods}. They are characterised by the following set of prescriptions: {\sf Matt+15} for the stellar wind torque, {\sf TZ97} for $\nu_v$ , and {\sf MPZ04} for $\nu_h$. We preferred the {\sf Matt+15} prescription for the torque over the formulation of {\sf Matt+12} , even though the latter directly takes into account magnetic field and mass-loss rate, precisely for sake of the consistency. Those are multi-dimensional processes that are still implemented in 1D models with many uncertainties. We show in Sect. \ref{sec:models} that both fit the observed rotation periods, but we prefer a simpler approach that includes fewer uncertain processes to ensure that we will not miss any effects.
\begin{table}[h]
\caption{Parameters and assumptions of the different models computed in this study: MPZ04 refers to \cite{MPZ2004}, TZ97 to \cite{TZ97},  Matt+15 to \cite{Mattetal2015}, Matt+12 to \cite{Matt2012}, R\&M 12 to \cite{RM12}, and GB13 to \cite{GB13}. SB is for solid-body rotation.}
\begin{center}
\begin{tabular}{ c | c | c | c c }
\hline
\hline
Initial P$_{rot}$, & $\nu_h$ & $\nu_v$&  Braking law & Ref. \\
DL time & prescrip. & prescrip. &  & \\
\hline
\hline
7 days & MPZ04 & TZ97 & Matt+15 & 1s \\ 
5 Myr & MPZ04 & Maeder97 &  Matt+15 & 2s \\
& Zahn92 & TZ97 &  Matt+15 & 3s \\
& Zahn92 & Maeder97 &  Matt+15 & 4s \\
& MPZ04 & TZ97 &  R\&M 12 & 5s \\
& MPZ04 & TZ97 &  Matt+12 & 6s \\
& MPZ04 & TZ97 &  K88 & 7s \\
& SB & SB &  Matt+15 & 8s \\
& MPZ04 & TZ97 &  GB13 & 9s \\
\hline
3.7 days & MPZ04 & TZ97 & Matt+15 & 1m \\ 
5 Myr &   MPZ04 & TZ97 &  R\&M 12 & 5m \\
& MPZ04 & TZ97 &  Matt+12 & 6m \\
& MPZ04 & TZ97 &  K88 & 7m \\
& SB & SB &  Matt+15 & 8m \\
& MPZ04 & TZ97 &  GB13 & 9m \\
\hline
1.4 days & MPZ04 & TZ97 & Matt+15 & 1f \\ 
3 Myr &  MPZ04 & Maeder97 &  Matt+15 & 2f \\
& Zahn92 & TZ97 &  Matt+15 & 3f \\
& Zahn92 & Maeder97 &  Matt+15 & 4f \\
& MPZ04 & TZ97 &  R\&M 12 & 5f \\
& MPZ04 & TZ97 &  Matt+12 & 6f \\
& MPZ04 & TZ97 &  K88 & 7f \\
& SB & SB &  Matt+15 & 8f \\
& MPZ04 & TZ97 &  GB13 & 9f \\
\hline
\end{tabular}
\label{tab:mods}
\end{center}
\end{table}

\subsection{Evolution of surface angular velocities}

Figure~\ref{Fig:refprot} shows the surface angular velocity evolution (in terms of rotation period) as a function of time for the three reference models ($1s$, $1f,$ and $1m$) as solid lines. 
Depending on their initial velocity, these models present very different behaviours, but they are qualitatively similar to those obtained by \citet{KPBS97}. The slow and medium rotators (models $1s$ and $1m$) experience a stronger stellar wind braking related to their initial angular momentum during the early evolution than do the fast rotator (model $1f$) towards the age of the present Sun. We also observe a convergence of the averaged rotation rate of the radiative interiors (dashed lines); this result is similar to previous work done with a slightly different treatment of AM transport \citep[e.g.][]{P90}. Consequently, all the models reach the age of the Sun with the same total amount of angular momentum independently of their initial content. We detail the evolution of the radiative region angular velocity profile to understand which processes are involved.

\begin{figure}[t]
\includegraphics[angle=270,width=0.47\textwidth]{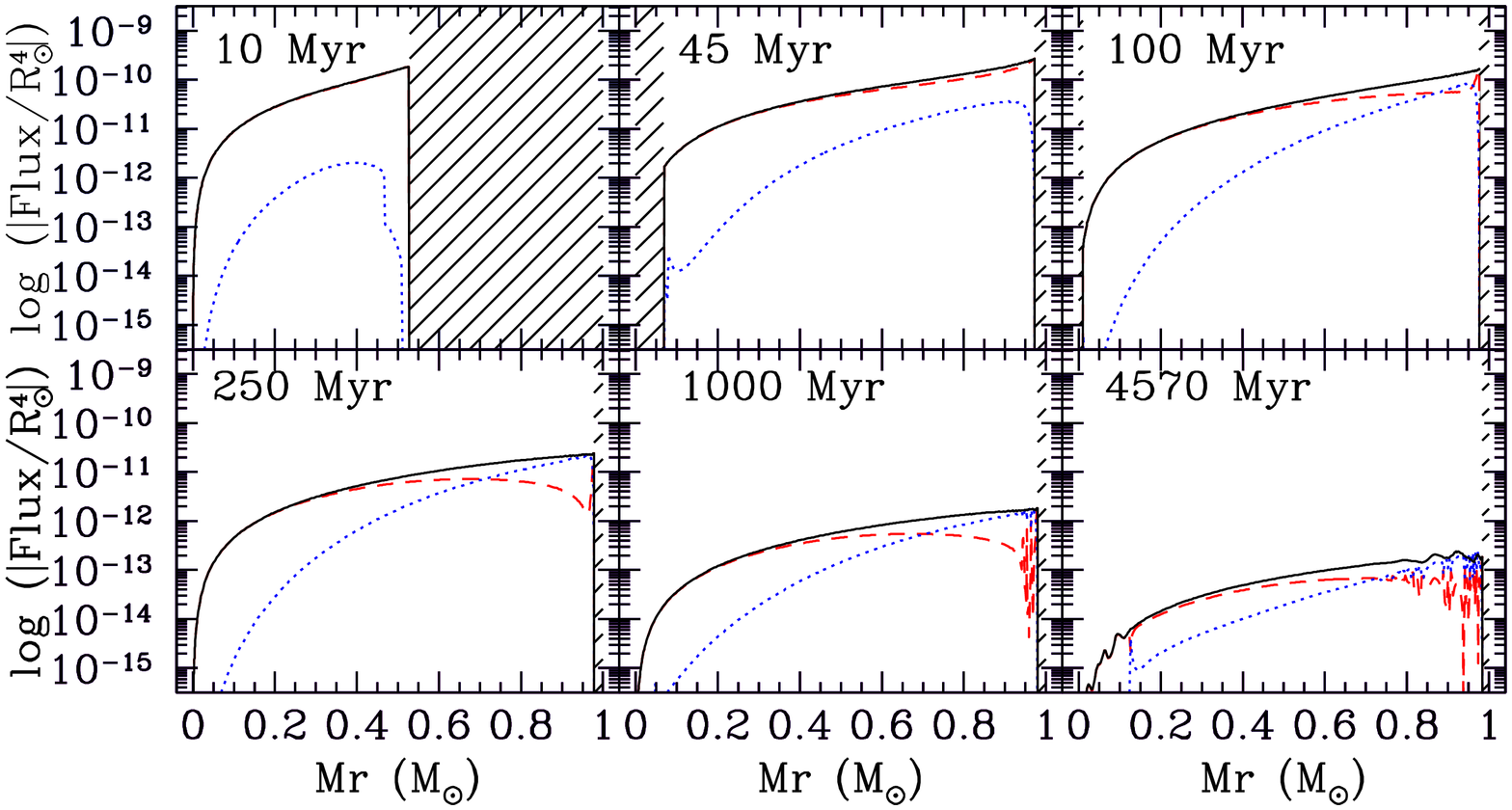}
\includegraphics[angle=270,width=0.47\textwidth]{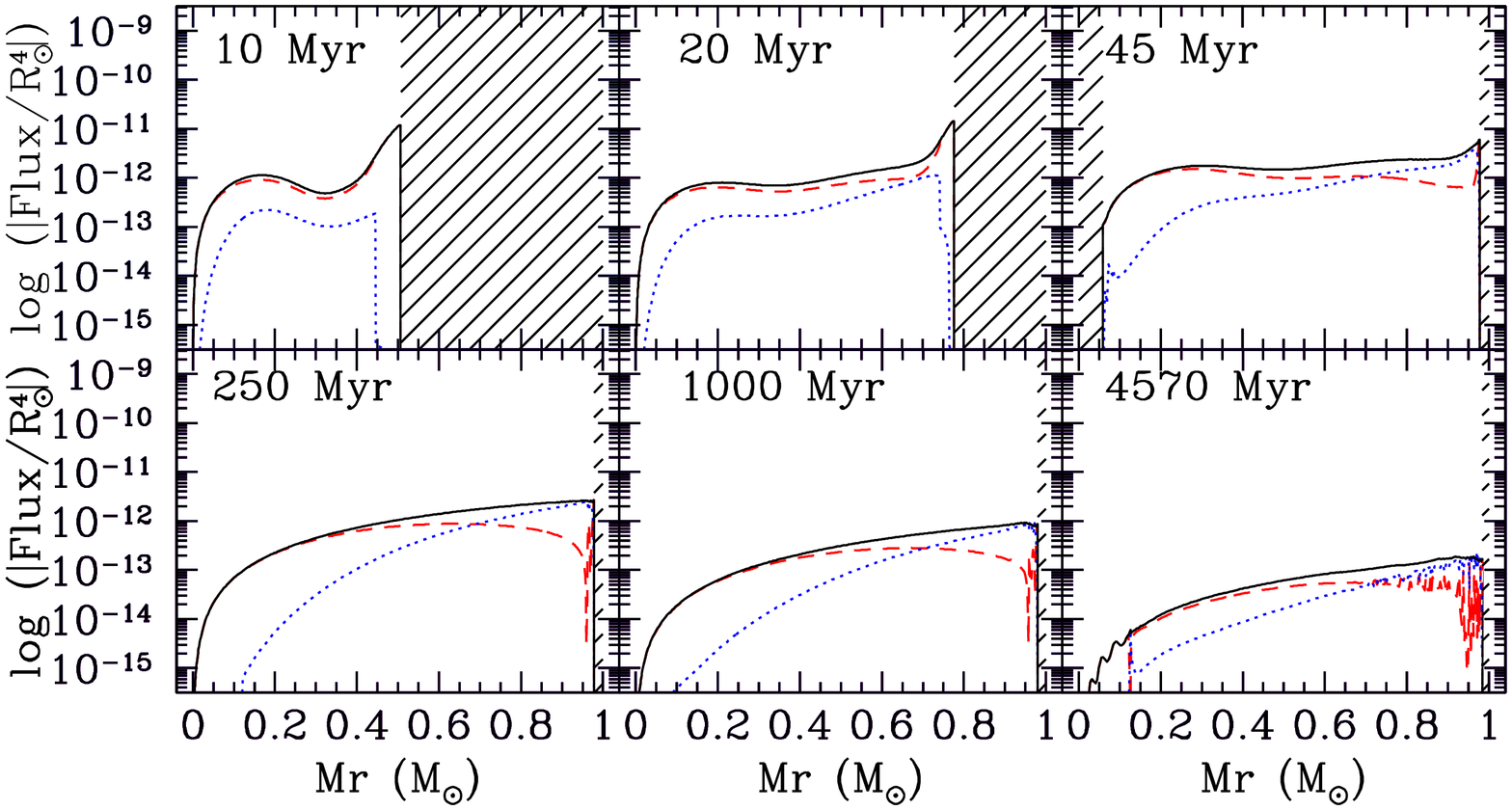}%
\caption{Angular momentum fluxes carried by meridional circulation (dashed red), shear (dotted blue), and the total flux (solid black) in the fast and slowly rotating models (top and bottom, respectively). Hatched areas in all plots indicate the convective regions.}
\label{Fig:reffluxAM}
\end{figure}

\subsection{Rotation profile and differential rotation}
Figure \ref{Fig:refprofomega} shows the evolution of the internal angular velocity profiles as a function of the mass coordinate for our reference models $1s$ ({\em left}), $1m$ ({\em central}), and $1f$ ({\em right}). We observe that for the special combination of braking and transport prescriptions adopted, all the models end with the same rotation profile (long-dash-dotted cyan profile), regardless of their initial angular momentum content. In all three cases, the differential rotation first increases (the core spins up) during the PMS evolution up to 45 Myr due to the contraction of the stars as they evolve along the Henyey track. Beyond 45 Myr, the core spins down in all models as a result of the meridional circulation that extracts AM from the core to compensate for the loss generated at the surface by the torque (see right panels of Fig.~\ref{Fig:reffluxAM} and below). The main difference between the three cases is the efficiency of the meridional circulation, which adjusts to the intensity of the torque: the larger the torque (for the fast rotators), the larger the angular velocity gradient and the more efficient the meridional circulation.
Moreover, since the efficiency of these angular momentum transport processes is proportional to the surface angular velocity, we observe a weak coupling between the radiative core and the envelope in models $1s$ and $1m$ at the same epochs (radiative interior rotating on average much faster than the convective envelope; see solid and dashed lines). Thus, the surface velocity evolution is already shaped by the torque before the stars reach the ZAMS (at $\approx$ 60 Myr), leading to their early spin down.\\
In contrast, the fast rotating model strongly couples the radiative and convective zone as long as the star contracts (spins up), and the internal angular momentum is carried outward. During these early phases, model $1f$ is in the saturated dynamo regime. At its arrival on the ZAMS, the star stops spinning up and the surface is strongly braked on a relatively short timescale. The surface velocity is divided by a factor of 10 in 200 Myr, until the star spins down enough to enter the unsaturated-dynamo regime and further decreases by another factor of 10 for the following 4.3 Gyr.

For clusters older than the Hyades (beyond $\approx$ 700 Myr), the observed distributions of rotation periods strongly narrow. 

By the time the models reach 1 Gyr, the torque has become much less efficient and the angular velocity profile of the fast rotator is similar to that of the median and slow rotators. The three models later follow the same evolution, which is dictated by the meridional circulation and shear turbulence until the age of the Sun. The surface angular velocity and the total angular momentum content of the slow, medium, and fast rotators are similar beyond 1 Gyr. The angular velocity profiles also share a similar shape and evolution in all three cases.

\begin{figure}
\includegraphics[width=0.45\textwidth]{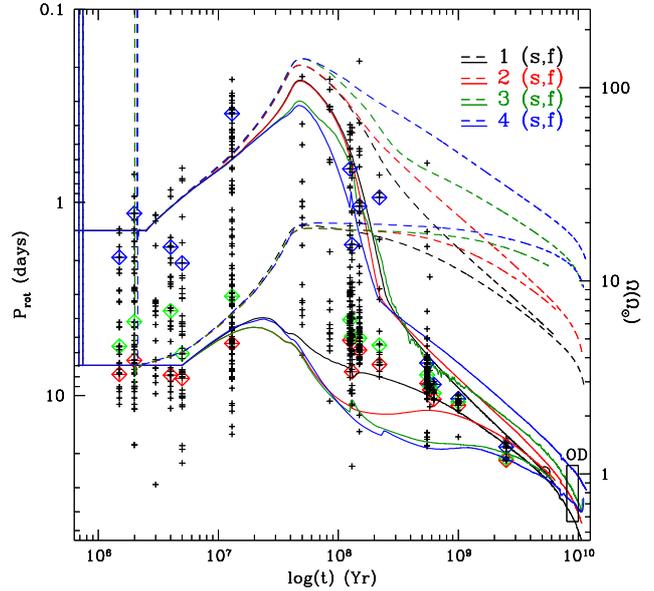}
\caption{Same as Fig.~\ref{Fig:refprot} for fast and slow rotator models computed with four different transport prescriptions. Among the slow rotators models $1s$, $2s$, $3s$, and $4s$ correspond to the tracks from top to bottom for an age of 500 Myr. The fast rotating models $1f$, $2f$, $3f$, and $4f$ follow he same order for an age of 60 Myr. The description of the physics used in each of these models is given in Table~\ref{tab:mods}.}
\label{Fig:protmixing}
\end{figure}

\subsection{Transport of AM}
Here we examine the processes that drive the internal transport of AM in our reference models. To do so, we followed the diagnostic procedure proposed by \citet{DecressinMathis2009}, which relies on the analysis of the AM fluxes associated with each transport process that is accounted for. The transport of angular momentum is mainly driven by the meridional circulation as long as the star is rotating relatively fast (above the saturation value), as shown in the left panels of Fig. \ref{Fig:reffluxAM}, and this independently of the initial velocity. After the ZAMS ($\approx$ 60 Myr), in all cases meridional circulation weakens because the star does not contract any longer and the surface velocity has already decreased. Angular velocity gradient increases at the base of the convective envelope, leading to a predominance of the shear turbulence over the meridional circulation in this region. For the slower rotators, the shear turbulence takes over the transport of AM in the outer radiative zone below the envelope and efficiently extracts (acting along the angular velocity gradient) the angular momentum, 
ensuring a continuous decrease of the surface angular velocity from the ZAMS until the age of the Hyades ($\approx$ 700 Myr) (see lower panel in Fig.~\ref{Fig:reffluxAM}). On the other hand, in the fast rotator, the meridional circulation is very efficient as long as the star  contracts (2 orders of magnitude larger than in the slow rotator) and efficiently couples the interior to the convective envelope up to the ZAMS. Beyond this age, its efficiency decreases rapidly, and the shear generated by the torque dominates below the convective envelope, although it is not sufficient to maintain an efficient transfer of AM from the core to the envelope.\\
The overall transport of AM by these processes tends to become less efficient as the star evolves on the main sequence as a
result of the decrease of the angular velocity gradient and of the torque.

Around the age of the Sun, some very tight reversals of the circulation that are due to the turbulence at very slow rotation rate appear close to the convective envelope bottom. The earlier evolution is dominated by an outward loop carrying angular momentum from the core to the surface to compensate for the loss that
is due to stellar wind.

\begin{figure}[t]
\includegraphics[width=0.45\textwidth]{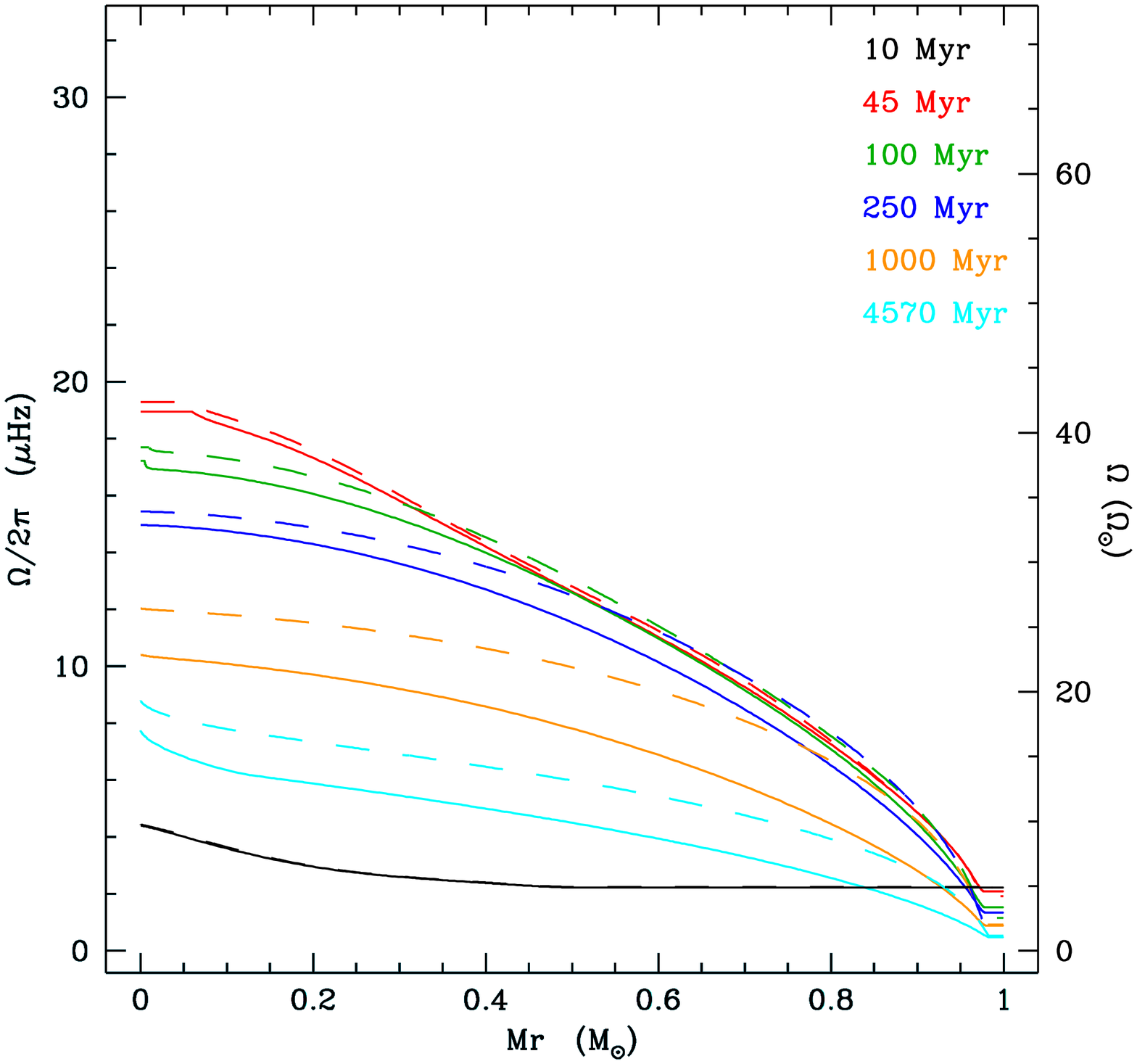}
\includegraphics[width=0.45\textwidth]{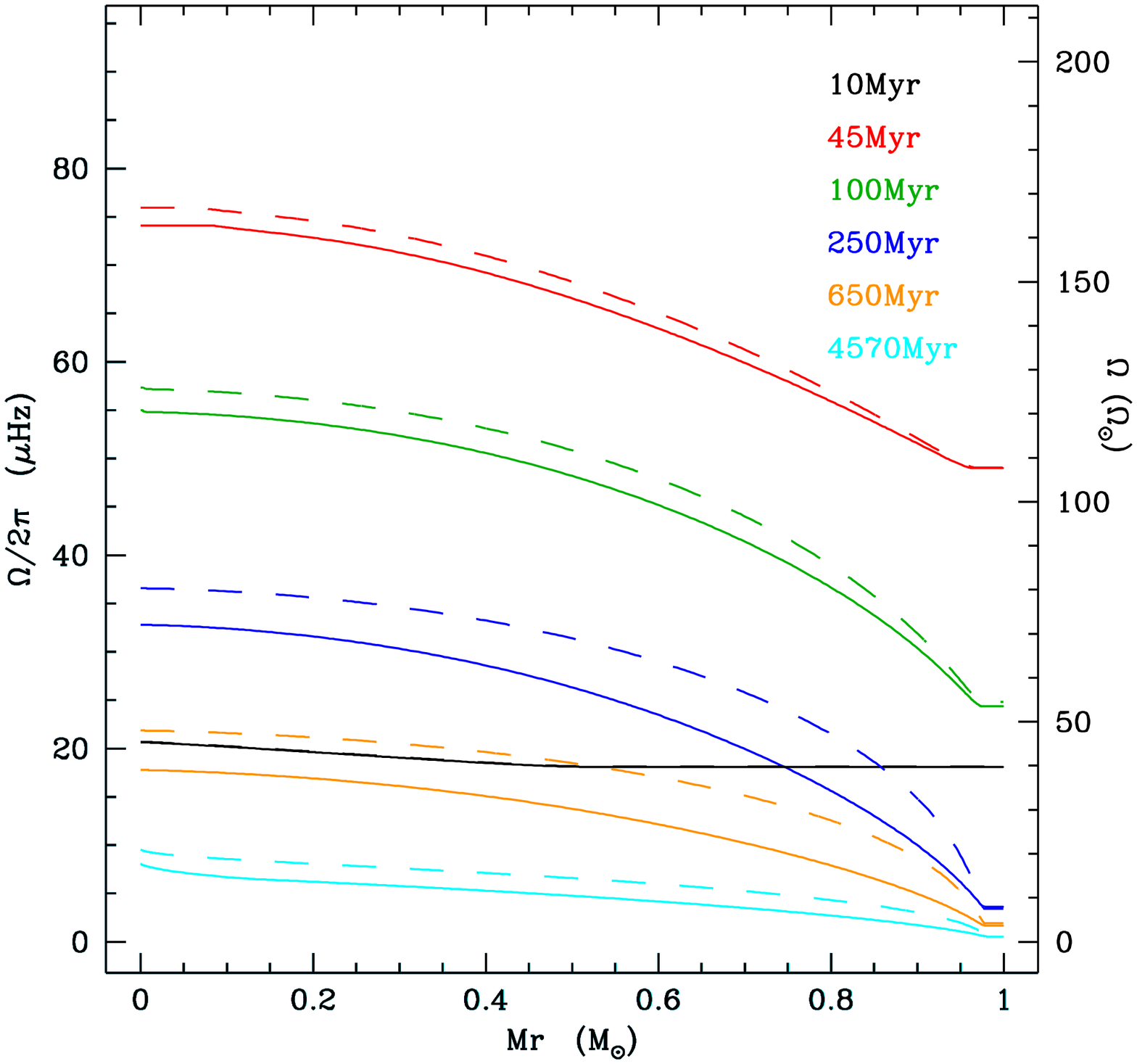}
\caption{Effect of modelling the vertical turbulent shear diffusion coefficient $D_v$ on the angular velocity profiles. Same as Fig.~\ref{Fig:refprofomega} for models $2s$ (dashed lines) and $1s$ (solid lines) in the top, and for models $2f$ (dashed lines) and $1f$ (solid lines) in the bottom panels. The description of the physics used in each of these models is given in Table~\ref{tab:mods}.}
\label{Fig:profomegaMaeder}
\end{figure}

\section{Exploring the physical parameter space}\label{sec:explore}

The description of our reference models points out the different roles played by the meridional circulation, the shear turbulence, and the torque in shaping the internal angular velocity profiles throughout the evolution. As indicated in Table~\ref{tab:mods} and Sect.~\ref{sec:torques}, several prescriptions exist to account for the transport of AM in the radiative interiors and to describe the stellar wind torque exerted at the surface. In this section, we analyse the effect of the prescription choice on the angular velocity and momentum evolution in solar-mass models. To do so, we use the set of diagnostic tools developed by \cite{DecressinMathis2009}.\\

\begin{figure}[t]
\includegraphics[width=0.45\textwidth]{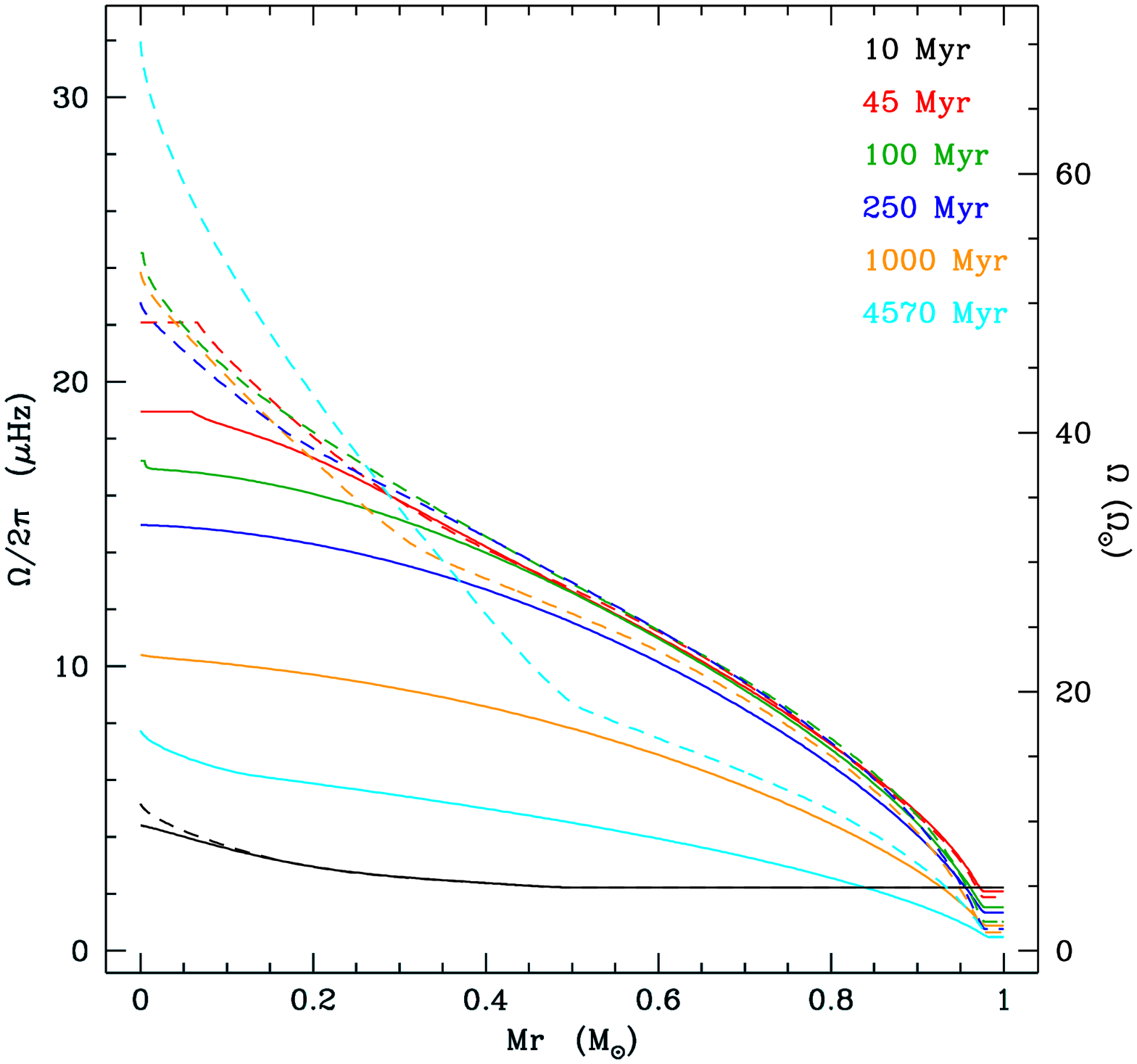}
\includegraphics[width=0.45\textwidth]{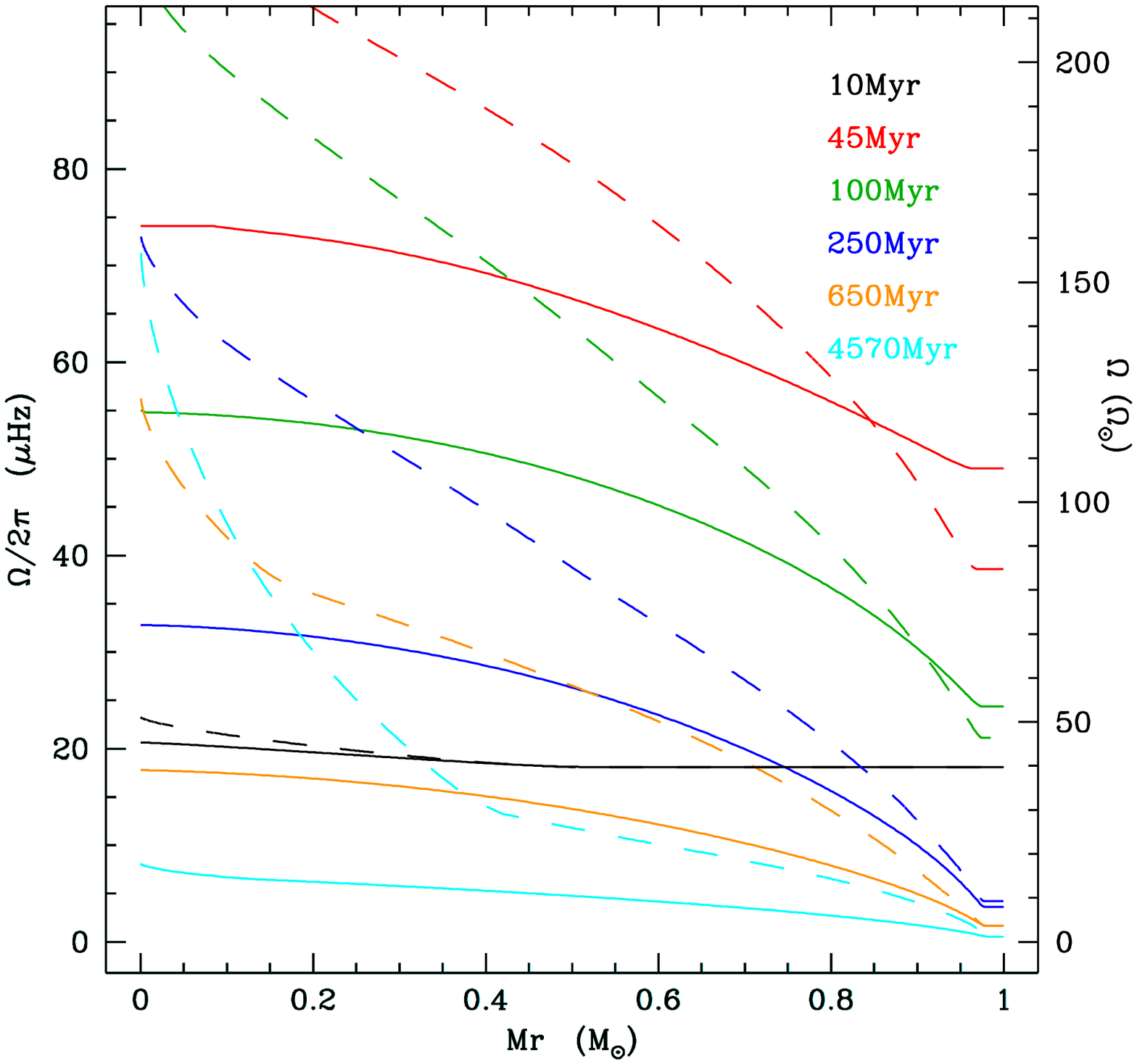}
\caption{Effect of the modelling of the horizontal turbulent shear diffusion coefficient $D_h$ on the angular velocity profiles. Same as Fig.~\ref{Fig:refprofomega} for models $3s$ (dashed lines) and $1s$ (solid lines) in the top, and for models $3f$ (dashed lines) and $1f$ (solid lines) in the bottom panels. The description of the physics used in each of these models is given in Table~\ref{tab:mods}.}
\label{Fig:profomegaZahn}
\end{figure}

\subsection{Transport of angular momentum in the stellar interior}

Within the paradigm described by \cite{Zahn1992} for the transport of angular momentum in radiative regions, several prescriptions have been developed to describe the shear turbulent viscosity in both the horizontal and vertical directions. While different groups prefer different combinations according to the type of stars studied, it was not until the study by \citet{Meynet2013} that the actual impact of choosing a specific combination of prescriptions on the evolution (structural, chemical, and rotational) was explored. These authors modelled massive stars ($9-40$M$_\odot$) and showed large differences in the HR diagram, the evolution
of surface chemical abundances, and the rotational evolution. However, the available observational data did not allow them to favour one of the combinations.

Here we analyse the effect of two different prescriptions for both $\nu_h$ and for $\nu_v$ on the rotational evolution of a 1 ~\Ms~ model. The overall effect on the surface and internal averaged angular velocity is shown in Fig.~\ref{Fig:protmixing}.\\

\subsection{Effect of the prescription for vertical shear}
We first compare models with the {\sf MPZ04} prescription for $D_h$ but with different prescriptions for the vertical shear ({\sf TZ97} or {\sf Maeder97}). The use of the {\sf Maeder97} prescription (given in Table~\ref{tab:coeffdiff}) for $\nu_v$ hinders the development of the vertical turbulent shear instability in a larger portion of the radiative interior, and the shear only contributes very little to the transport of AM throughout the evolution. For the slow rotator (model $2s$ in Fig.~\ref{Fig:protmixing} and top panel in Fig.~\ref{Fig:profomegaMaeder}), this leads to a sharp decrease of the surface angular velocity from the ZAMS after the contraction of the star has stopped and the meridional circulation is weaker. Between 100 Myr and  1 Gyr, the efficiency of the meridional circulation to transport AM remains almost constant in the inner radiative zone and ensures the transport of AM from the core to the envelope to compensate for the extraction of AM at the surface by the stellar wind torque. Therefore this leads to an almost constant surface angular velocity over this period of time.

For the fast rotators (model $2f$ in Fig.~\ref{Fig:protmixing} and lower panel in Fig.~\ref{Fig:profomegaMaeder}), on the other hand, the prescription of $\nu_v$ has a very weak influence on the surface velocity because the transport is always dominated by the meridional circulation, which is of similar amplitude as in the reference model. The weak effect of the vertical shear diffusion prescription on the AM evolution also appears when looking at the evolution of the total specific angular momentum as a function of time shown in Fig.~\ref{Fig:momspec} (dotted line, left panel).\\

\subsection{Effect of the prescription for horizontal turbulence}
The effect of the choice of the prescription for the horizontal shear turbulence is much more obvious: in Fig.~\ref{Fig:protmixing} the two models with $\nu_h$ from {\sf Zahn92} (models $3s, 3f, 4s,$ and $4f$
) present a very high ratio $\Omega_{interior}/\Omega_{surface}$ compared to the models $1s$ (resp. $1f$) and $2s$ (resp. $2f$) computed with $\nu_h$ from {\sf MPZ04} (see Table~\ref{tab:mods}). 

Using the expression of {\sf Zahn92}  for $\nu_h$ leads to a meridional circulation that is overall slower than for $\nu_h$ of {\sf MPZ04}. 

\begin{figure}
\includegraphics[width=0.48\textwidth]{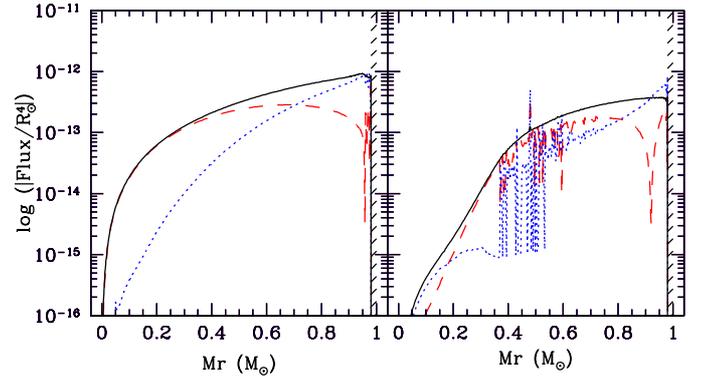}
\caption{ Same legend as Fig.~\ref{Fig:reffluxAM} for two slow rotators at 1 Gyr computed with $\nu_h$ from {\sf MPZ04} (left) and {\sf Zahn92} (right).}
\label{Fig:2dcirc_Dh}
\end{figure}

 The turbulence can develop and disturb the meridional circulation in the radiative interior if the Richardson criterion is fulfilled. In our reference model, the circulation is strong enough to keep its original shape even in presence of shear turbulence, but this is not always the case. Thus, in models $3s$($f$) and $4s$($f$), using $\nu_{h,Zahn92}$, the angular velocity profiles are much steeper, as shown in Fig.~\ref{Fig:profomegaZahn}, and trigger a much stronger mean shear-induced turbulence. The loop that was carrying AM outward then breaks into several reverse cells as soon as the Richardson criterion is fulfilled (shown in Fig.~\ref{Fig:2dcirc_Dh}) and annihilates AM transport locally. Consequently, a break appears in the angular velocity profile that can be seen in Fig.~\ref{Fig:profomegaZahn} around 0.3 M$_r$ (resp. 0.2 M$_r$) at 1 Gyr and around 0.5 M$_r$ (resp. 0.4 M$_r$) at the solar age for the slow (resp. fast) rotator.

In addition, the surface velocity at the ZAMS and during the early evolution on the main sequence is lower than in the reference models, leading to a less efficient torque (which is directly proportional to the surface angular velocity) and to a global larger amount of specific angular momentum when the star reaches the age of the Sun (see Fig.~\ref{Fig:momspec}).

\subsection{Stellar wind torque}

We now explore the effect of the stellar wind torque prescriptions on the overall angular momentum and angular velocity evolution of solar-mass stars.\\ First, we calibrated each of the prescriptions detailed in Sect.~\ref{sec:torques} to reproduce the surface angular velocity of the present Sun at 4.57 Gyr. To do so, we calibrated one parameter, $\mathcal{C}, K_1$ and $K_W$ for breaking laws of {\sf R\&M12, Matt+12,} and {\sf K88}, respectively. Second, we adjusted a second parameter\footnote{$\Omega_{\rm sat}$, $m$ and $\Omega_{\rm sat}$ for ${\rm R\&M12}$, ${\rm Matt+12,}$ and ${\rm K88, respectively.}$} to adjust the data in younger clusters, if possible. The obtained values are given in Table~\ref{tab:brake}, and for every model with a given braking law, the same set of parameters was used, regardless of the initial angular momentum content. \\


\begin{figure*}[t]
\includegraphics[width=0.95\textwidth]{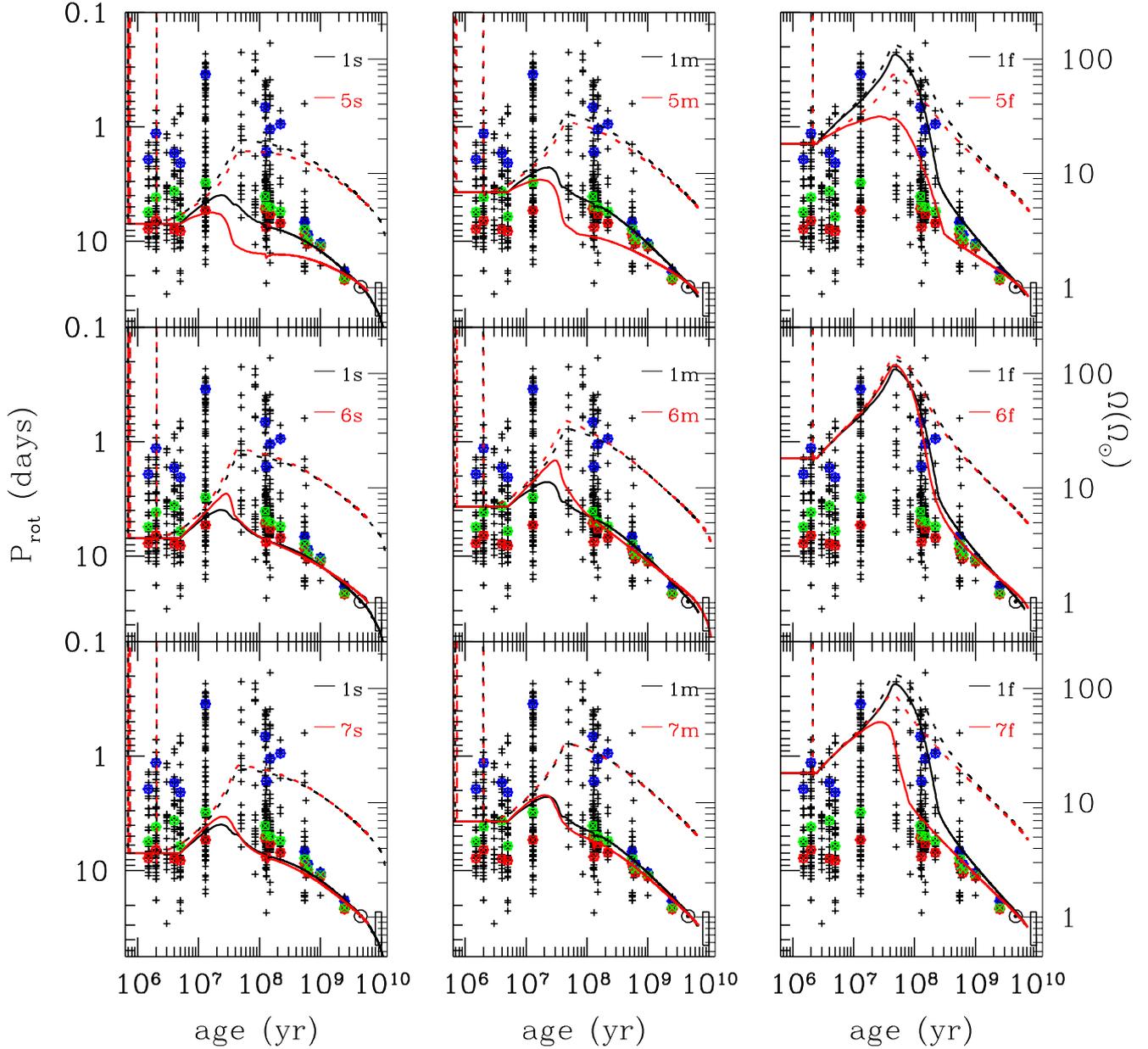}
\caption{Same legend as Fig. \ref{Fig:refprot} with different braking laws than in our reference model. The references of the models are given in Table~\ref{tab:mods}. The reference model corresponds to the black tracks.}
\label{Fig:prot}
\end{figure*}

Figure \ref{Fig:prot} presents the evolution of the surface rotation for slow, medium, and fast rotators  with different torque prescriptions, but with the same internal AM transport description ({\sf MPZ04, TZ97}). 

\begin{figure}
\includegraphics[width=0.45\textwidth]{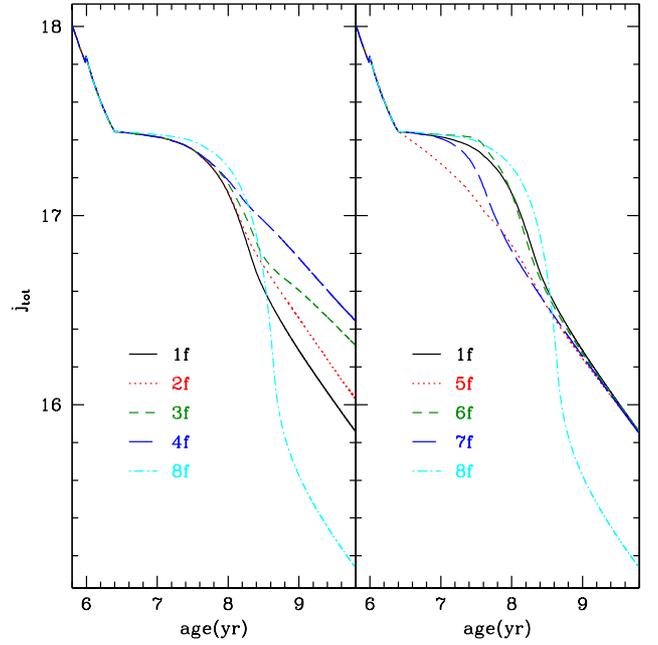}
\caption{Evolution of the total specific angular momentum as a function of time. ({\em Left}) Effect of the AM transport prescriptions. ({\em Right}) Effect of the torque prescriptions.}
\label{Fig:momspec}
\end{figure}

We note that in all cases, the mean angular velocity of the radiative core reaches the same rotation period, regardless of the stellar wind torque prescription used, as shown in Fig. \ref{Fig:momspec}.

\subsubsection*{\sf RM12}
The surface rotation evolution using this prescription is compared to the reference models in the upper row of Fig. \ref{Fig:prot}. The strong radius dependency of the breaking law reported in
\citet{RM12} explains the very high loss of angular momentum observed in Fig. \ref{Fig:couple} ($5f$ track in the upper right panel) during the PMS, when the star is still contracting.  The angular velocity increases very slowly during the PMS compared to the reference model, and sharply drops at the ZAMS. As soon as the velocity drops below the saturation value ($\Omega_{sat,RM12} = 3\Omega_\odot$), the extraction of angular momentum almost stops and the surface angular velocity evolution stabilises to converge to $\Omega(t) \propto t^{-0.25}$ (see Eq. \ref{eq:RM12}).

\subsubsection*{\sf Matt+12}
The surface rotation evolution using this prescription is compared to the reference models in the middle row of Fig. \ref{Fig:prot}. The angular velocity evolution of fast rotators is the same as in the reference model, despite the absence of explicit saturation value in {\sf Matt+12} prescription. It appears in the BOREAS routine as an asymptotic regime of the magnetic field filling factor $f_\star$ that is perfectly mimicked by the saturation coefficient $\chi$ we took for the reference model using {\sf Matt+15} prescription. 
A difference appears around the ZAMS for slow rotators, and because of this difference of saturation, models with the torque according
to {\sf Matt+12}  reach higher rotation periods. Nevertheless, the discrepancy remains very small and as we lack data for these ages, there is no real constraint to determine which one is better.

\subsubsection*{\sf K88}
\begin{figure}[t]
\includegraphics[width=0.45\textwidth]{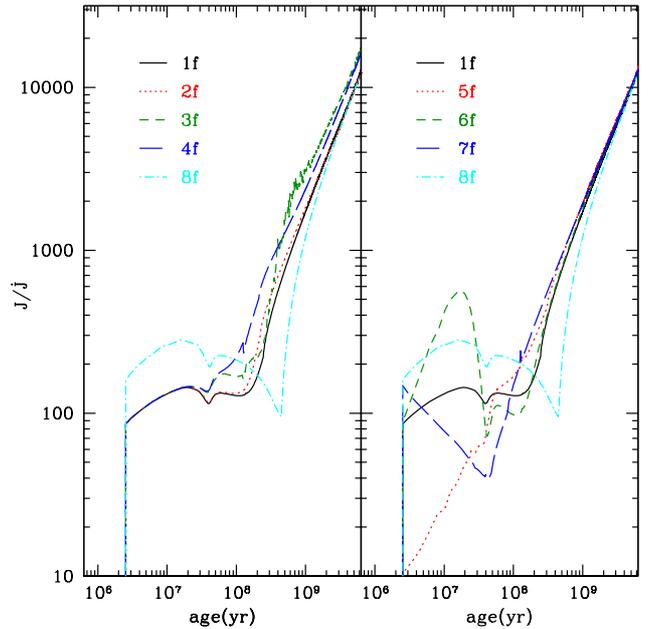}
\caption{Same as Fig.~\ref{Fig:momspec} for the evolution of the ratio between the total angular momentum and the angular momentum loss due to wind alone (disc-coupling loss not accounted for).}
\label{Fig:couple}
\end{figure}

The surface rotation evolution using this prescription is compared to the reference models in the lower row of Fig. \ref{Fig:prot}. Both slow and medium rotators braked with {\sf K88} behave as the reference model in the unsaturated regime. Still, to obtain a solar surface angular velocity at the age of the Sun, we need to take a high $K_W$ value. The extraction of angular momentum is therefore three times higher at the ZAMS than for the reference model (see right panel of Fig. \ref{Fig:couple}).
This strong braking prevents the models from achieving rapid rotation at the age of the Pleiades with the {\sf K88} torque. This result is similar to what was obtained in previous work \citep[e.g.][]{Pinsonneaultetal1989}.

\section{Comparison to bi-zone models}\label{sec:bizone}


In this section we compare our predictions with those of the bi-zone models of \citet{GB13}.
Their approach splits the star into two regions, each of them rotating as a solid body. The external region corresponds to the convective envelope, the inner region to the inside layers of the star (including the radiative region and, if exists, the convective core).
Exchange of AM between these two zones is assumed to occur at a given rate along the whole evolution.
The empirical coupling timescale is constrained to reproduce the observed surface rotation periods.

For a comparison between the two approaches, we deduced a coupling timescale between the convective envelope and the rest of the star for our full evolutionary models.  This is possible thanks to the determination of the angular momentum flux transferred by meridional circulation and shear turbulence between the external convective region and the radiative part of the star (see Eq. \ref{eq:fluxeq} in Sect. \ref{sec:formalism}).

This flux was then introduced into the formulation of \cite{MB1991} to determine a timescale,
\begin{equation}
\tau_c = \Delta J \times \frac{1}{F_{\rm tot}}
,\end{equation}
where 
\begin{equation}
\Delta J = \frac{I_{\rm conv}J_{\rm rad} -I_{\rm rad}J_{\rm conv}}{I_{\rm rad} + I_{\rm conv}}
,\end{equation}
with $I_{\rm conv}$, $J_{\rm conv}$ and $I_{\rm rad}$, $J_{\rm rad}$ being the inertial and angular momentum of the convective envelope and radiative zone (and if exists, the convective core).

Because the amount of AM that is transported is mainly dominated by the meridional circulation, its corresponding timescale is two orders of magnitude shorter than the one associated with diffusive turbulence. The timescales associated with both meridional circulation and shear turbulence are much longer than the constant coupling time-scale used in bi-zone models \citep[e.g.][]{GB13,GB15}.
These models are calibrated to reproduce the open cluster observations and the surface velocity, the total angular momentum, and the mass-loss rate of the present Sun. The assumed coupling does not necessarily depend on time, rotation rate, or differential rotation rate.

To properly compare our models to the bi-zone models of \citet{GB13,GB15}, we computed a slowly rotating model with exactly the same braking law as described in Table 2 of \citet{GB13} (model {\em 9s}), and a bi-zone model with the same parameters for the torque and for the coupling timescale as in \citet{GB13}, but based on the structural evolution of our STAREVOL models (model {\em GB13})\footnote{In \citet{GB13}, the evolution of the structural quantities, e.g. M, R, inertia momenta, are taken from \citet{Baraffe98} grid. }. 

\begin{figure*}[t]
\includegraphics[width=0.95\textwidth]{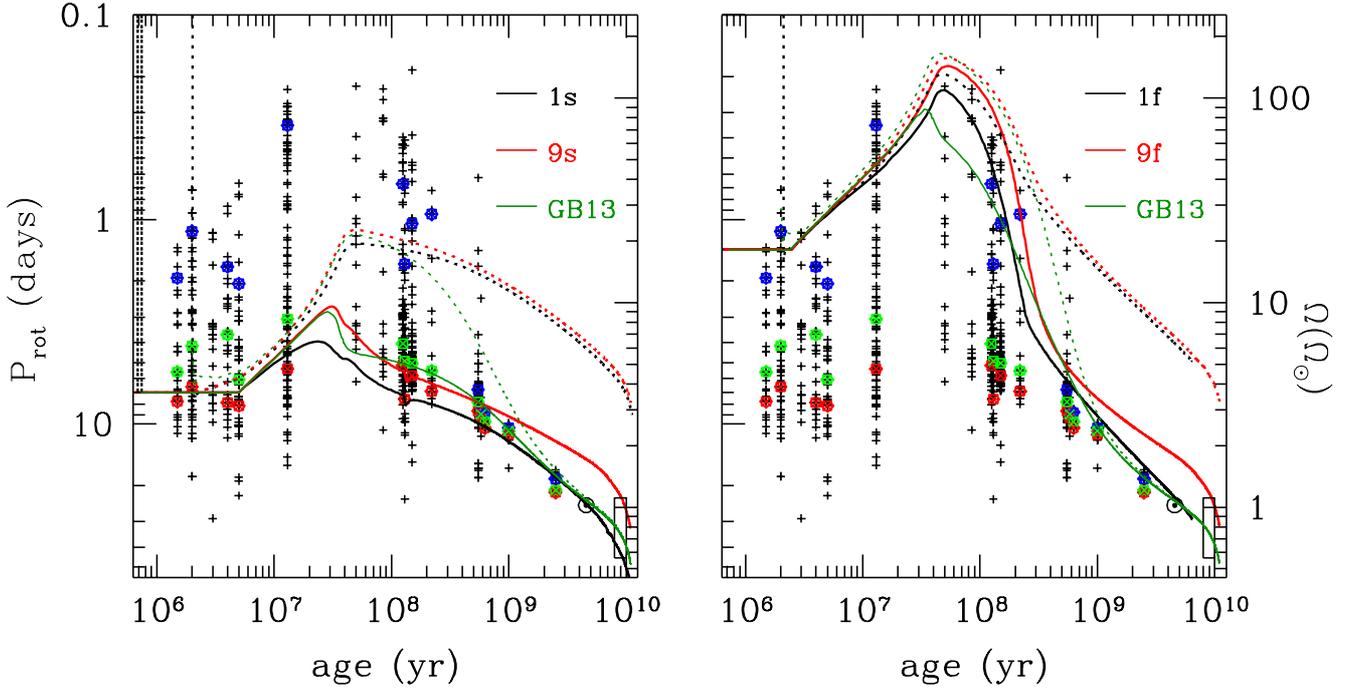}

\caption{Surface rotation evolution as a function of time in the reference slow ({\em left}) and fast ({\em right}) models compared to bi-zone models ({\sf GB13}). Data points are as described in Fig. \ref{Fig:refprot}. The references of the models are given in Table~\ref{tab:mods}.}
\label{Fig:prot2}
\end{figure*}

\begin{figure}
\includegraphics[width=0.42\textwidth]{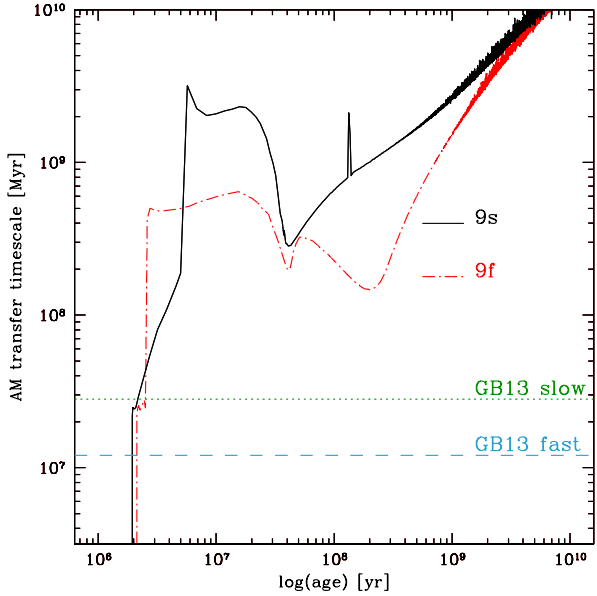}
\caption{Coupling timescale from the AM exchange between the radiative core and the convective envelope.}
\label{Fig:Timescales}
\end{figure}

Thus, the only differences between the two models are the transport of angular momentum in the radiative core and the exchange with the convective region. The evolution of the surface velocity of these models as a function of time is shown in Fig.~\ref{Fig:prot2}.
Model $9s$ solves a self-consistent time-dependent AM exchange between the radiative interior and the convective envelope, with an evolving associated timescale, while the $GB13$ model assumes a constant transfer of AM between the interior and the convective envelope, leading to a constant timescale throughout the evolution, as shown in Fig. \ref{Fig:Timescales}. 
This can qualitatively be transcribed as a forced transfer of AM from the radiative to the convective region that changes the surface velocity and therefore the angular velocity-dependent AM extraction at the surface, while in our self-consistent model the AM extraction triggers the AM transport below the convective region and thus changes the core-envelope coupling.

At the very beginning of the evolution, when the star is still coupled to its disc but is already partially radiative, the forcing of the disc on the stellar surface leads to a very strong meridional circulation, and the corresponding AM transport timescale is very short compared to the rest of the evolution.
After the star decouples from the disc, the forcing stops and the AM transfer timescale increases sharply. Beyond this point, the coupling timescale in model $9s$(f) is always much longer than the timescale assumed in the bi-zone model $GB13$ (fast or slow). On the MS the structure of the star stabilizes and the core-envelope coupling follows both the evolution of the surface angular velocity and the differential rotation. 
We can note that the fast rotators experience a much stronger coupling during the whole evolution. This is especially evident around 200 Myr when the star is still a fast rotator and the differential rotation quickly increases because of the strong braking that slows down the surface. Since both the absolute and the differential rotation rate play a role in the efficiency of AM transport, the coupling timescale reaches a minimum. Even though the convective and radiative regions of the fast rotator are more strongly coupled than for the slow rotator, the coupling timescale remains more than one order of magnitude higher than the constant coupling timescale set for the two-zones model.

\begin{figure}
\includegraphics[width=0.42\textwidth]{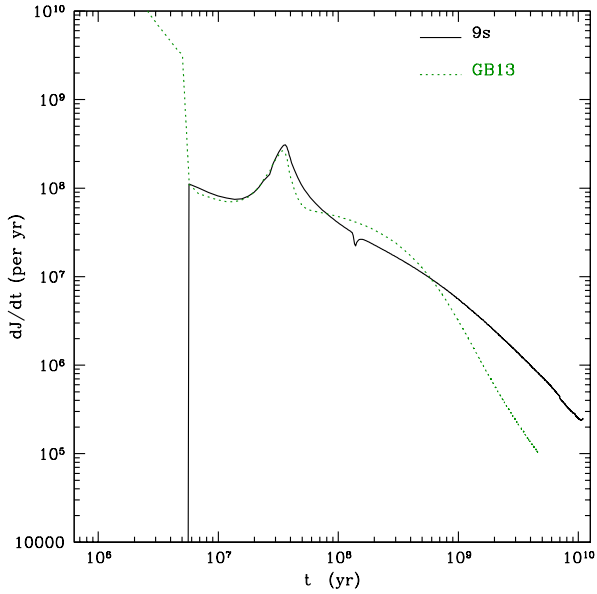}
\caption{Specific angular momentum loss per year as a function of time.}
\label{Fig:TorqueGB13}
\end{figure}

Figure \ref{Fig:TorqueGB13} shows the extraction of angular momentum by magnetized winds in both cases. As expected, the evolution is similar during the PMS because the evolution of angular velocity is driven by structural effects that are identical in both models. After the star stops contracting, the internal angular momentum transport becomes relevant and a difference appears between the two models. As more AM is brought to the surface, the angular velocity is higher with a stronger coupling in model $GB13$. When the angular velocity is higher for the same structure, the extraction by the winds is stronger during this phase and pumps out most of the radiative core AM reservoir. All the AM brought to the convective region is immediately removed by stellar winds
because a certain equilibrium is reached between the core-envelope coupling and the surface extraction.
Around 1 Gyr, the model with $GB13$ parameters  rotates almost as a solid body, the transfer of AM from the core to the external envelope becomes very weak and causes the radiative core to rotate at the same angular velocity as the convective region. At the same time,  model $9s$ loses AM faster, but the rate is still two orders of magnitude lower than at the ZAMS. All the AM loss that matters for the AM content at the age of the Sun occurs before 1 Gyr.

\section{Discussion with respect to additional constraints}\label{sec:other}

\subsection{Constraint on internal rotation from helioseismology}
\begin{figure}
\includegraphics[width=0.45\textwidth]{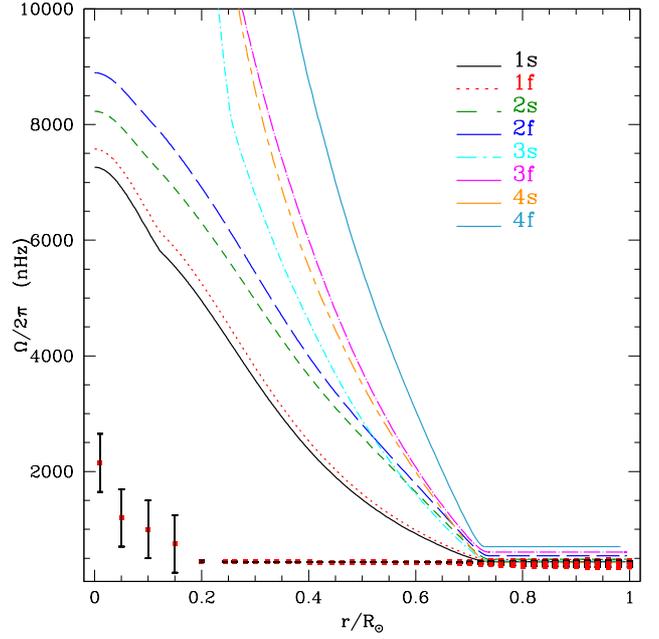}
\caption{Comparison of our models with the rotation profile of the Sun obtained from helioseismology by \citet{Garcia2011}}
\label{Fig:Sunprof}
\end{figure}

Asteroseismology, and more specifically helioseismology, has
delivered very accurate data that allow probing the internal structure of stars. \citet{ThompsonScience1996} showed that the solar rotation profile can be inverted deep in the radiative zone, down to about 0.2 $R_\odot$, using the information carried by the rotational splitting of high-order low-degree $p$-modes.
More recently, and as shown in Fig. \ref{Fig:Sunprof}, \citet{Garcia2011} used rotational splitting of candidate $g$-modes to deduce the rotation profile of the Sun in its very core, below $0.2 R_\odot$.
The $p$-mode analysis shows that the Sun is rotating almost as a solid body in the radiative region down to about $0.2 R_\odot$, while candidate g-modes seem to indicate an increase of the angular velocity from $\simeq 430$ nHz to 2 $\mu$Hz in the innermost regions. This latest result is still debated, however. We here neglect a number of AM transport processes that are expected to modify the angular velocity profile in the radiative interior \citep{CharbonnelTalon2005Science,Eggenberger2005,Charbonnel2013}, so that we do not expect our models to fit the helioseismic profile.
When plotted against helioseismic data, all the models discussed previously are, as expected, far from observations in the radiative region. Indeed, even the model with the flatter rotation profile that we obtain at the age of the Sun still spins four times faster in the central region and does not reproduce the solid-body region between 0.2 R$_\odot$ and the bottom of the convective envelope
at all.
Internal gravity waves \citep{CharbonnelTalon2005Science} and magnetic fields \citep{Eggenberger2005} have shown very promising results in reproducing the solar rotation profile. However, these processes appear to act on very different timescales, and incorporating
them is beyond the scope of the present study, where we focus on hydrodynamical processes.

\subsection{Constraint from lithium abundances}
\begin{figure*}
\includegraphics[width = 0.45\textwidth]{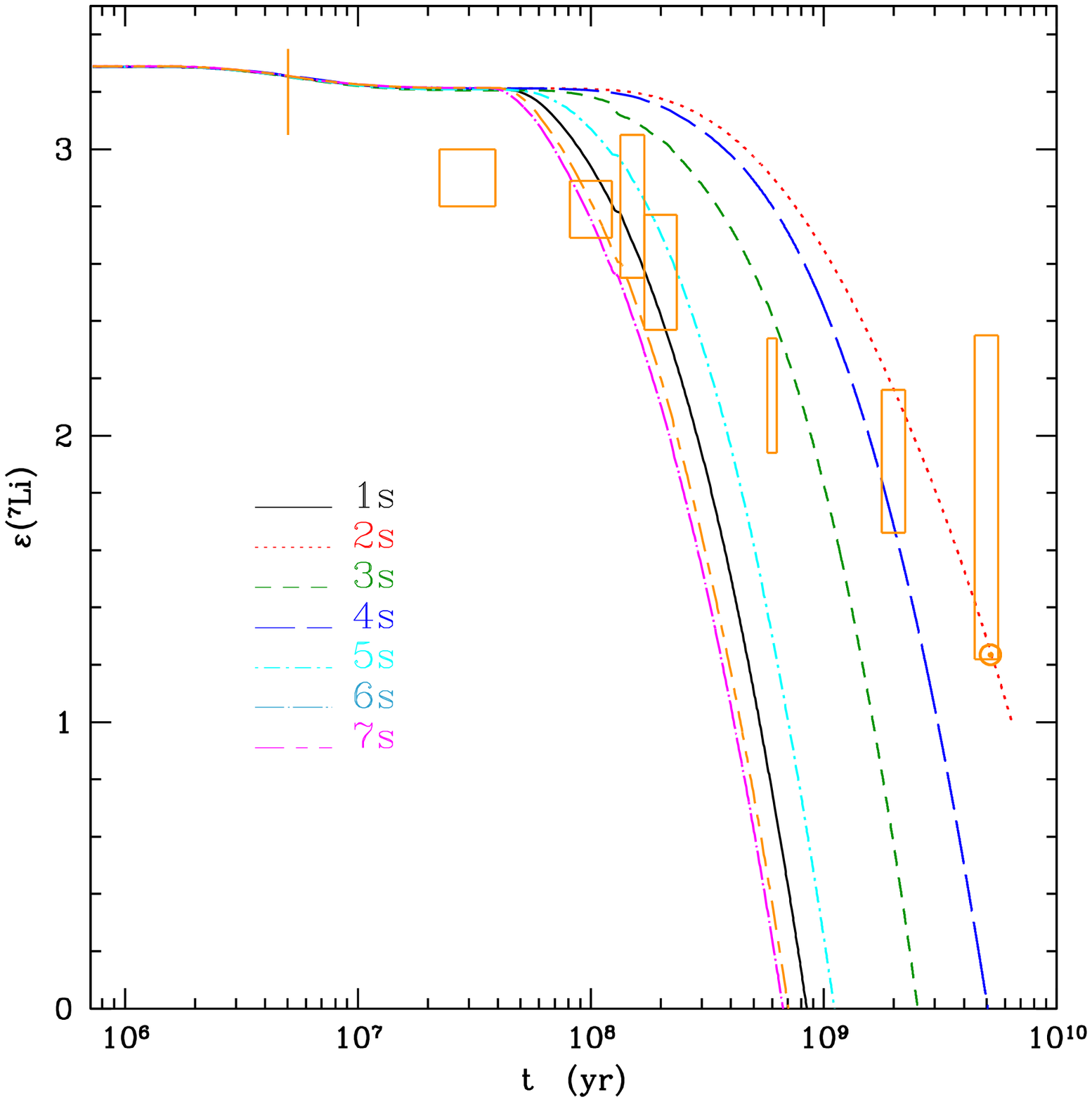}%
\includegraphics[width = 0.45\textwidth]{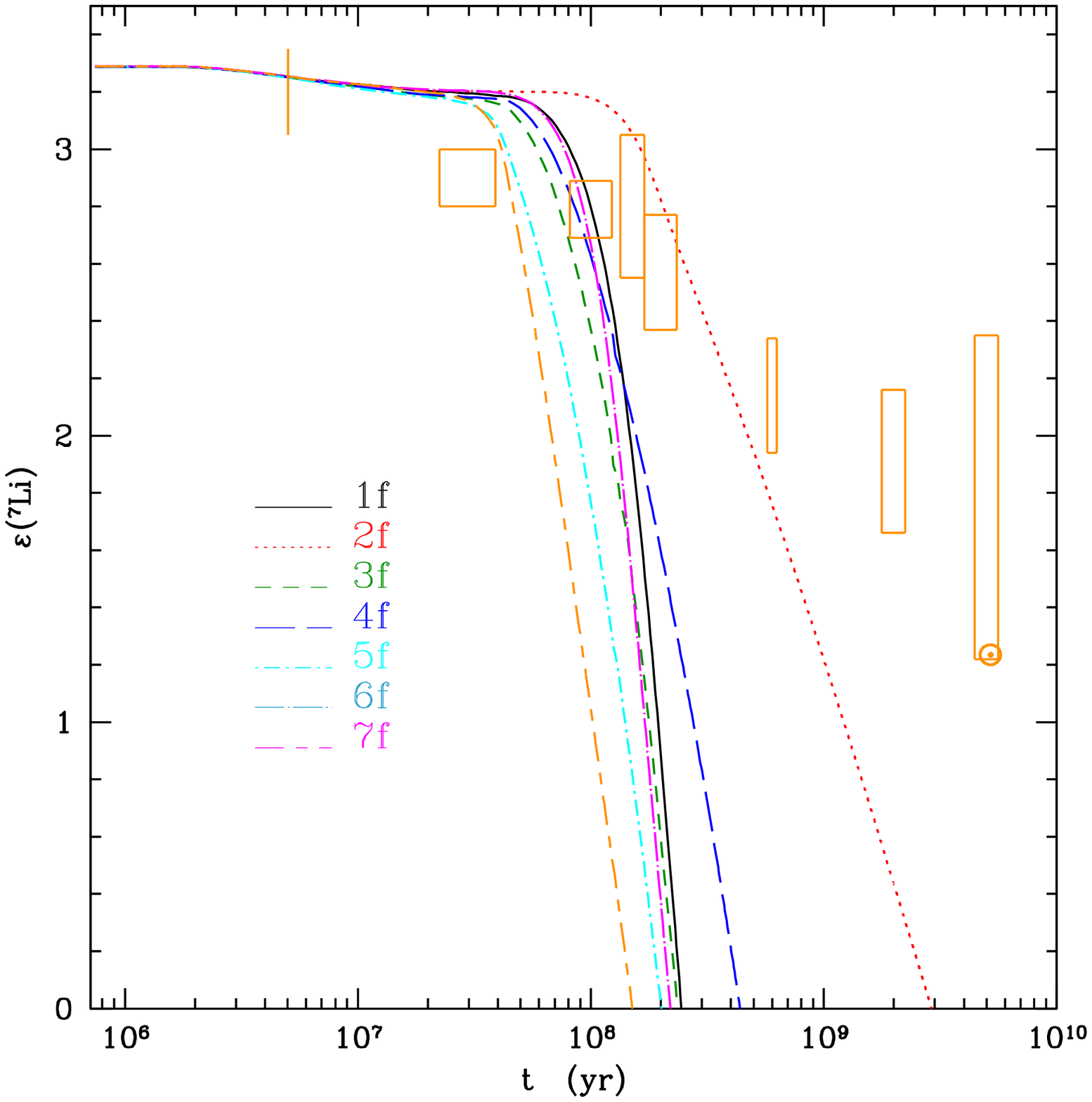}
\caption{Evolution of the surface lithium abundances in the slow ({\em left}) and fast ({\em right}) rotating models as a function of time. The open boxes represent the ranges of lithium abundances determined by \citet{SestitoRandich2005} for the following clusters and associations: NGC2264 at 5 Myr, IC2391/IC2602/IC4665 at 30 Myr, Pleiades/Blanco1 at 80 Myr, NGC2516 at 150 Myr, M34/M35/NGC6475 at 250 Myr, Hyades/Praesepe/Coma Ber/NGC6633 at 600 Myr, NGC752/NGC3680/IC4651 at 2 Gyr and M67 at 4.5 Gyr. The Sun is also indicated with the solar symbol $\odot$.}
\label{Fig:Li}
\end{figure*}

Because lithium is prone to  destruction by proton captures at temperatures $T \geq 2.5 \times 10^6$ K, its abundance at the stellar surface is a good proxy to evaluate the depth and efficiency of mixing processes that may connect the low-temperature convective envelopes of stars to deeper radiative regions where lithium is destroyed \citep[e.g.][]{Deliyannisetal00}.
Considering that meridional circulation and turbulent shear instability not only transport AM but also matter, it is therefore interesting to look at the predictions of our models concerning the evolution of the lithium surface abundance. It worth noting that in our self-consistent approach, the lithium depletion predicted by the models is a direct consequence of the evolution angular rotation profile driven by the extraction and internal transport of AM (i.e. we do not have any adjustable parameter to reduce the efficiency of the mixing to fit the Li data). \\ The results are shown in Fig.~\ref{Fig:Li}, where we compare the predictions of the models to the observed range of Li abundances determined for solar-type stars in different open clusters from 5 Myr to the age of the Sun as compiled by \citet{SestitoRandich2005}\footnote{To establish these ranges, we selected the stars in each cluster that have the temperature of a solar-type star of the same age according to stellar evolution models.}. As expected from previous works, our reference models (both the slow and the fast rotators) selected to best reproduce the angular velocity evolution of solar-type stars fail to simultaneously reproduce the observed lithium abundance evolution of solar-type stars on the main sequence (beyond 250 Myr (resp. 100 Myr) for the slow (resp. fast) rotators). This is due to the efficient transport of lithium by the shear turbulent instability in the radiative interior of our models. The lithium destruction occurs earlier in the fast rotators because of the correlation between the shear mixing efficiency and the angular velocity gradient, which is directly related to the torque. \citet{CharbonnelTalon2005Science} obtained similar results and showed that the introduction of internal gravity waves could reconcile the models predictions with observations by flattening out the angular velocity profile and thus reducing the efficiency of the turbulent shear mixing.\\
We point out the unexpected behaviour of model {\em 2s} that is computed using \citet{Maeder1997} prescription for the vertical turbulent shear diffusivity. In this case, the effective diffusion coefficient representing the transport of nuclides by the meridional circulation is the same as in our reference case {\em 1s} because we used the same prescription for the $D_h$ expression. However, $D_v$ from {\sf Maeder97} is significantly lower than that predicted by {\sf TZ97}, in particular in regions where the mean molecular weight gradients are large. As a result, the transport of nuclides is less efficient in model {\em 2s} and the surface abundance of lithium decreases much more slowly than in our reference model {\em 1s}.
 In summary, none of the present models that account for the rotational behaviour of solar-type stars is able to simultaneously account for the lithium evolution and the helioseismic constraints, as already anticipated in previous works of our group.

\section{Summary and global picture}\label{sec:end}
Observations have shown that the AM of solar-mass stars is reduced by two orders of magnitude during the first few Myr of their interaction with their disc. After this phase, they need to decrease it by two more orders of magnitude to reach the solar angular momentum content. This loss of AM is thought to occur through stellar wind coupled with the large-scale stellar magnetic field. However, since young stars are not expected to rotate as solid bodies, the transport of angular momentum in the stellar radiative interiors should modify the evolution of their surface velocity at every age. Hydrodynamical processes transporting AM are driven by the forcing induced by the extraction of AM at the surface by stellar winds. Thus, depending on the wind-braking prescription used in the models, the efficiency of the transport will be different.  

Different prescriptions for the AM extraction and transport processes exist in the literature, with different degrees of approximation. 
We here showed that self-consistent 1 M$_\odot$ rotating evolutionary models that account for the transport of AM by meridional circulation and shear turbulence alone and that use a realistic AM wind-driven extraction can reproduce the evolution of the surface rotation of open clusters from the disc-coupling phase to the age of the Sun. Using different prescription sets for the horizontal and vertical diffusion coefficients essentially impacts the AM distribution inside the star. 

The surface extraction of AM and the AM transport in the radiative region interact in a complex way, the first triggering the later and the later enforcing the first.\\
We find that the stellar wind prescriptions from \citet{Matt2012,Mattetal2015} provide a really good match with observations of rotation periods in open clusters when they are combined with the adapted AM transport, meaning here {\sf MPZ04} and {\sf TZ97} for the prescriptions of the horizontal diffusion and vertical diffusion coefficients, respectively. None of the other braking laws lead to as good results, independently of the chosen prescriptions for internal AM transport. 

This led us to select the following combination for our reference model: ($D_v$; $D_h$; $dJ/dt$) $\equiv$ ({\sf TZ97, MPZ04, Matt15}). With this set of prescriptions, our models all reached the same internal rotation profile at the age of the Sun, regardless of the assumed initial angular momentum content. 

Unlike in bi-zone models, the angular momentum content of the present Sun cannot be retrieved with our models. Even using the most favourable combination of the available prescriptions, meridional circulation and shear-induced turbulence are therefore
just efficient enough to reproduce the observed $P_{rot}$ data, but additional processes are still needed to reproduce asteroseismic and surface chemical abundances observations. Regardless of the set of AM transport prescriptions we used, we were unable to reproduce these observed data.

In the recent years, the asteroseismic data mostly gathered by the \textit{Kepler} mission for subgiant and red giant stars also revealed the need to go beyond the current description of rotational mixing to be able to account for the rotational properties (surface and core rotation) of evolved low-mass stars \citep{Cellieretal2012,VSP2013,Garciaetal2014,Deheuvelsetal2014,Deheuvelsetal2015}. 
The dominant processes that shape the AM evolution may differ according to the evolutionary phase. However, observational evidence now exists, for stars ranging from the PMS to the red clump, pointing towards the need of additional AM transport mechanisms that would efficiently couple the core and the envelope of solar-type stars.

As predicted by \citet{MacGregor2000} and tested in main-sequence evolutionary models by \citet{CharbonnelTalon2005Science} and \citet{Eggenberger2005}, magnetic field and internal gravity waves are very good candidates to provide additional angular momentum transport and should be investigated in a further study.\\ It is also important to investigate whether our conclusions also apply to other low-mass stars in the mass range 0.5~\Ms~ to 2~\Ms~. The structural and rotational evolution is expected to differ depending on the initial mass in this mass range \citep[see e.g.][]{GB15}, which may have a significant effect on the efficiency of possible transport processes, as was discussed for instance by \citet{TalonCharbonnel2003}.

\begin{acknowledgements}
This study was supported by the grant ANR
2011 Blanc SIMI5-6 020 01 “Toupies: Towards understanding the spin
evolution of stars” (http:\/\/ipag.osug.fr\/Anr\_Toupies\/ ). 
C. Charbonnel and F. Gallet acknowledge support from the European COST Action TD 1308 Origins and the SEFRI project C14.0049 of the Conf\'ed\'eration Suisse.
\end{acknowledgements}

\bibliography{BibADS}




\appendix
\section{Magnetic field generation and mass-loss rate}
Magnetic field and mass loss were computed following \cite{CS2011}. Nevertheless, to calibrate our model on the Sun, we needed to modify some expression in the BOREAS routine that were rewritten in Fortran. These are set out in this appendix.

The effective temperature was directly taken from STAREVOL and
was not computed from luminosity and radius as in the initial routine.

Following \citet{GB13}, we reduced the kink wave energy flux $F_A$ (Eq. (7) in \citet{CS2011}) by a factor $2.5$ to obtain a mass-loss rate of $2.9 \times 10^{-14}  M_\odot.yr^{-1}$ at the age of the Sun, which is still consistent with the estimated range of the actual solar mass-loss rate ($2 - 3.2\times 10^{-14} M_\odot.yr^{-1}$). 

Similarly to \citet{GB13}, we took the same expression as $f_{min}$ in \citet{CS2011}, but slightly modified it to reproduce the average filling factor of the present Sun ($f_{*\odot} = 0.0035 \in [0.001;0.01]$):
\begin{equation}
f_\star = \frac{0.4}{[1+\left(x/0.16\right)^{2.3}]^{1.22}},
\end{equation}
where $x$ is the normalized Rossby number. $\tau_{conv}$ and $\tau_{conv_\odot}$ are determined with the expression given in the BOREAS routine. It is a function of the effective temperature, which was fitted to a set of ZAMS models given by \citet{Gunn98}. 

\end{document}